\documentclass[aps,prl,showpacs,twocolumn,superscriptaddress]{revtex4-2}

\usepackage{times,xspace}
\usepackage{graphicx,graphics,color,epsfig}
\usepackage{amsmath}
\usepackage{amssymb}
\usepackage{amsfonts}
\usepackage{amsfonts}
\usepackage{epstopdf}
\usepackage{bm}
\usepackage{hyperref}
\usepackage{color}
\usepackage{float}
\usepackage[normalem]{ulem}
\usepackage{xcolor}

\def\be{\begin{equation}}
\def\ee{\end{equation}}
\def\ba{\begin{eqnarray}}
\def\ea{\end{eqnarray}}

\begin{document}

\title{High-temperature \texorpdfstring{$\eta$}{eta}-pairing superconductivity in the photodoped Hubbard model}

\author{Lei Geng}
\email{lei.geng@unifr.ch}
\affiliation{Department of Physics, University of Fribourg, Fribourg-1700, Switzerland}
\author{Aaram J. Kim}
\affiliation{Department of Physics and Chemistry, DGIST, Daegu 42988, Korea}
\author{Philipp Werner}
\email{philipp.werner@unifr.ch}
\affiliation{Department of Physics, University of Fribourg, Fribourg-1700, Switzerland}

\date{\today}
 
\begin{abstract}
We investigate superconductivity emerging in 
the photodoped Mott insulating Hubbard model
using steady-state dynamical mean-field theory implemented on the real-frequency axis. By employing high-order strong-coupling impurity solvers, we obtain 
the nonequilibrium 
phase diagram for photoinduced $\eta$-pairing superconductivity with a remarkably high effective critical temperature. We further identify a superconducting gap in the momentum-resolved spectral function and optical conductivity, providing spectroscopic signatures accessible to experiments. Our results highlight a route to a controllable form of high-temperature superconductivity in nonequilibrium strongly correlated systems, fundamentally distinct from the equilibrium $s$-wave pairing state in the attractive Hubbard model or cuprate-like $d$-wave superconductors.
\end{abstract}
\vspace{0.5in}

\maketitle

Superconductivity, characterized by zero electrical resistance and the expulsion of magnetic fields, is a central research topic in condensed matter physics, due to both its striking quantum nature and potential for technological applications~\cite{bardeen1957theory,tinkham2004introduction}. Experimentally, a wide variety of superconducting materials have been discovered over the past decades~\cite{Takabayashi2016,stewart2011superconductivity,RevModPhys.78.17,RevModPhys.75.473,kamihara2008iron,sun2023signatures}. Despite this diversity, transition temperatures at ambient pressure remain far below room temperature, and practical applications for high-temperature superconductors are very limited. 
Although various microscopic mechanisms—such as spin fluctuations~\cite{PhysRevLett.17.433,Werner2016}, resonating valence bonds~\cite{anderson1987resonating}, and  polarons~\cite{alexandrov2010advances,alexandrov1981bipolaronic}—have been proposed~\cite{RevModPhys.84.1383,RevModPhys.78.17}, the complexity of correlated electron materials has so far prevented a detailed understanding, making theoretical predictions of high-$T_c$ superconductivity in these systems highly challenging.

Recently, alternative nonequilibrium routes to superconductivity have attracted considerable interest. Experimentally, signatures of superconducting-like behavior have been reported in laser-driven correlated materials~\cite{Kaiser2014,cavalleri2018photo,budden2021evidence,rowe2023resonant,isoyama2021light}. 
On the theoretical side, Yang showed that the half-filled Hubbard model supports a family of exact eigenstates with off-diagonal long-range order, the so-called $\eta$-pairing states~\cite{yang1989}. These states correspond to Cooper pairs with finite center-of-mass momentum and exhibit transport properties analogous to conventional superconductors. In the strong-coupling Mott regime, such $\eta$-paired states become particularly relevant for nonequilibrium realizations. Photoexcitation across the Mott gap creates doublon--holon pairs, which can serve as the building blocks of $\eta$ pairs. Because doublon--holon recombination is strongly suppressed in large-gap systems \cite{Sensarma2010,Eckstein2011}, the density of doublon--holon pairs is approximately conserved, leading to long-lived steady states. Numerical studies based on exact diagonalization (ED) and density-matrix renormalization group (DMRG) methods have demonstrated that such photodoped Mott insulators in one dimension indeed support $\eta$-pairing superconductivity at zero temperature~\cite{kaneko2019photoinduced,murakami2022exploring}. Related behavior~\cite{Li2020} has also been reported for the infinite-dimensional Bethe lattice with (transient or permanent) coupling to external baths~\cite{Werner2019,li2021nonequilibrium}.

In contrast to these model studies, most experimentally relevant correlated materials are (quasi-)two-dimensional or three-dimensional. Accurate nonequilibrium simulations of such systems are numerically challenging, which has so far hindered a detailed assessment under which conditions $\eta$-pairing superconductivity can be realized in realistic settings. Even basic quantitative questions—such as the transition temperature as a function of photodoping, or the existence of a superconducting gap with clear spectroscopic signatures—remain unresolved, thereby impeding experimental efforts to identify and characterize this state.

In this work, we address these questions by studying the two- and three-dimensional 
Hubbard model in the photodoped Mott regime, using steady-state dynamical mean-field theory (DMFT) \cite{li2021nonequilibrium} combined with our recently developed third-order strong-coupling impurity solver~\cite{kim2025strong,Geng2025third}. 
We show that the $\eta$-pairing state exhibits effective transition temperatures comparable to those of the equilibrium attractive Hubbard model, far exceeding the $T_c$ of $d$-wave superconductors. 
It persists down to low photodoping levels $\delta$, with the effective transition temperature remaining above room temperature even at $\delta\!\approx\!0.1$. The spectral functions and the optical conductivity display clear superconducting gap signatures, providing experimentally accessible evidence of $\eta$ pairing.

\begin{figure}[t]{
    \centering
    \includegraphics[width=0.46\textwidth]{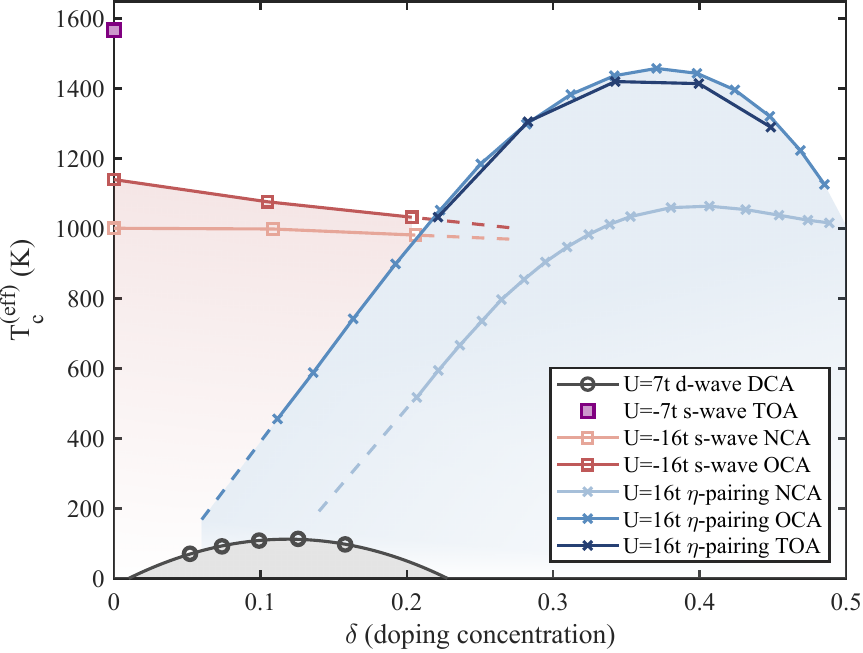}
\caption{
(Effective) superconducting transition temperatures versus carrier doping for chemically and photodoped Hubbard systems. 
Gray circles and line: $d$-wave transition temperatures of the repulsive Hubbard model at $U=7t$ from DCA calculations in Ref.~\cite{Dong2019}. 
Light- and dark-red squares: DMFT results for the attractive Hubbard model at $U=-16t$ from NCA and OCA; purple filled squares: TOA result for $U=-7t$. 
Competing charge order is suppressed. 
Crosses: photodoped Hubbard model solved within nonequilibrium DMFT using NCA (light blue), OCA (blue), and TOA (dark blue).
}
\label{fig_phasediagram}
}
\end{figure}

The Hubbard model is described by the Hamiltonian 
\begin{equation}
H = -t \sum_{\langle ij\rangle,\sigma}
\left( c_{i\sigma}^\dagger c_{j\sigma} + \mathrm{H.c.} \right)
+ U \sum_i n_{i\uparrow} n_{i\downarrow} - \mu \sum_{i\sigma} n_{i\sigma},
\end{equation}
where $t$ is the nearest-neighbor hopping, $U$ the on-site interaction, and $\mu=U/2$ the chemical potential. 
To solve the model, we employ DMFT, which maps the lattice problem onto a self-consistent Anderson impurity model~\cite{Georges1996,Aoki2014,Metzner1989}. The impurity problem is solved directly on the real-frequency axis using our quantics tensor cross interpolation (QTCI)~\cite{Ritter2024,Yuriel2025,Yue2023} strong-coupling solvers~\cite{Geng2025third,kim2025strong}. 
These solvers provide high-resolution spectral functions without analytic continuation and allow 
us to study the effects of 
higher-order diagrammatic corrections~\cite{Eckstein2010}.

Since we focus on superconducting phases with spontaneous $U(1)$ symmetry breaking, the DMFT equations are formulated in symmetry-adapted Nambu bases for both conventional and $\eta$-pairing states (see Supplemental Material (SM)~\cite{SM}, which includes Refs.~\cite{ginzburg2009theory,Nayak2025,Martin2008,Haule_2007,Ferrell1958,Tinkham1956}). A small pairing seed field 
is introduced to stabilize the broken-symmetry solutions, 
and the transition temperature is extracted from the anomalous expectation value $\langle c_{\downarrow}c_{\uparrow}\rangle$ using its mean-field square-root scaling $\propto \sqrt{T_c-T}$ \cite{SM}.

In this study, we are interested in long-lived photodoped Mott states~\cite{murakami2025photoinduced}, i.e., nonequilibrium steady states rather than thermal equilibrium phases. Consequently, standard imaginary-time equilibrium approaches are not applicable. Instead, we work on the real-frequency axis and impose the steady-state distribution by shifting the effective Fermi levels of the photocarriers (doublons/holons in the upper/lower Hubbard band) to $\pm\omega_F$, while fixing a common effective temperature $T^\text{eff}$~\cite{kunzel2024numerically}. This procedure enables 
the simulation of photodoped states without introducing auxiliary baths that may distort the spectra. 

The coexistence of strong local correlations and photodoped metallic carriers places stringent demands on the impurity solver. As demonstrated in our previous work~\cite{Geng2025third} and the SM~\cite{SM}, low-order strong-coupling schemes such as the non-crossing approximation (NCA) and one-crossing approximation (OCA) cannot accurately resolve the photodoped spectra. The third-order approximation (TOA) is more reliable and hence essential for the calculations presented in this study.

In the main text, we focus on the two-dimensional Hubbard model with $t=0.35\,\mathrm{eV}$ \footnote{This hopping parameter has been used in simulations of $d$-wave superconductivity~\cite{Dong2019}}. Results for the three-dimensional model can be found in the SM \cite{SM}. While DMFT lacks nonlocal correlations, it is meaningful to show data for the two-dimensional photodoped model, since the relevant temperature scales and carrier densities are high. 
We simulate the nonequilibrium system for different photodoping levels at $U=16t$. This large interaction is chosen to maintain a sizable Mott gap and thus a long-lived photodoped state under strong excitation. The photodoping concentration $\delta$ is defined as the electronic occupation above $\omega=0$.  

Figure~\ref{fig_phasediagram} plots the effective transition temperatures $T^\text{eff}_c$ for the $\eta$-pairing state, with increasingly dark shades of blue from NCA to TOA.
All approaches consistently predict remarkably high $T^\text{eff}_c$ over a broad range of photodoping, 
with a pronounced dome structure in OCA and TOA. Toward lower photodoping, $T^\text{eff}_c$ decreases approximately linearly with decreasing carrier concentration and it is expected to vanish as the photodoping concentration $\delta\rightarrow 0$. Within OCA the superconducting solution can be numerically stabilized down to at least $\delta\!\approx\!0.1$, where $T^\text{eff}_c$ is still above room temperature. Quantitatively, NCA substantially underestimates $T^\text{eff}_c$ compared to the higher-order solvers, while TOA closely follows the OCA results in the dome region. In the figure, we hence mark the $\eta$-pairing regime based on the OCA phase boundary. 

For comparison, we also include several equilibrium results, where the horizontal axis denotes chemical doping rather than photodoping. Gray symbols show dynamical cluster approximation (DCA) results for the repulsive Hubbard model at $U=7t$, taken from Ref.~\cite{Dong2019}, yielding a maximal $d$-wave $T_c$ of order $100\,\mathrm{K}$. This is within the expected range for cuprate superconductors~\cite{RevModPhys.75.473}, but far below the $T^\text{eff}_c$ obtained for the $\eta$-pairing state. We furthermore solved the attractive Hubbard model, which can be realized in cold atom systems \cite{Esslinger2010}, 
but not in conventional materials. Since competing charge order correlations are suppressed, the DMFT calculations yield robust $s$-wave superconductivity with high transition temperatures, exemplified by $T_c \sim 1100\,\mathrm{K}$ for $U=-16t$, and only a weak dependence on doping $\delta$ in the low-doping regime \footnote{In the exact phase diagram for the two-dimensional Hubbard model, $T_c\rightarrow 0$ for $\delta\rightarrow 0$ \cite{Fontenele2022}}. In addition, a TOA calculation at half filling yields an even higher transition temperature of $T_c \sim 1600\,\mathrm{K}$ for $U=-7t$.
Notably, these values are comparable to the transition temperatures we obtain in the strongly photodoped repulsive-$U$ model. This similarity can be understood as follows: while the attractive Hubbard model features an explicit on-site attraction that binds Cooper pairs, the strongly photodoped Mott insulator develops an effective $\eta$-pair attraction in its low-energy description. Indeed, a Schrieffer--Wolff transformation in the strong-coupling regime $U \gg t$ yields the effective low-energy model~\cite{Li2020}
\begin{equation}
H_{\mathrm{eff}}
= \frac{4t^{2}}{U}
\sum_{\langle ij\rangle}
\left(
\mathbf{S}_i \cdot \mathbf{S}_j
-
\boldsymbol{\eta}_i \cdot \boldsymbol{\eta}_j
\right),
\label{effective_Hamiltonian}
\end{equation}
where $\mathbf{S}_i=\frac{1}{2}c_{i\alpha}^\dagger \boldsymbol{\sigma}_{\alpha\beta} c_{i\beta}$ denotes the local spin operator and 
$\boldsymbol{\eta}_i$ is the pseudospin operator defined by
$\eta_i^+=(-1)^i c_{i\uparrow}^\dagger c_{i\downarrow}^\dagger$,
$\eta_i^-=(\eta_i^+)^\dagger$, and
$\eta_i^z=(n_i-1)/2$.
The ferromagnetic coupling in the $\eta$-spin sector favors staggered long-range $\eta$-pairing order at large photodoping within the leading-order strong-coupling description. However, the dome-shaped dependence of $T^\text{eff}_c$ with an optimal value away from the fully photodoped limit comes from higher-order corrections.

Figure~\ref{fig_phasediagram} shows that the $\eta$-pairing transition temperatures obtained with OCA and TOA are quantitatively very close, which raises the question of why the computationally costly inclusion of third-order diagrams is necessary. The differences between OCA and TOA primarily manifest themselves at the level of the Green's functions, even when the phase boundaries appear converged. The third order correction is particularly important in the superconducting state, where causality imposes a stringent constraint on the self-energy: the magnitude of the imaginary part of the anomalous component $\Sigma^\text{A}$ must not exceed that of the normal one $\Sigma^\text{N}$~\cite{SM},
\begin{equation}
\left| \mathrm{Im}\,\Sigma^{\mathrm{N}}(\omega) \right|
\;\ge\;
\left| \mathrm{Im}\,\Sigma^{\mathrm{A}}(\omega) \right| .
\label{causality}
\end{equation}
 NCA and OCA calculations typically violate this constraint already slightly below the transition temperature, resulting in numerical instabilities, whereas TOA provides access deep into the superconducting regime. Only at this level can the spectral properties of the $\eta$-pairing state, such as the gap structure and spectral-weight redistribution, be clearly resolved.

 This behavior is not known \emph{a priori}. Although perturbative expansions can be formally extended to arbitrarily high order and are generally asymptotic, their convergence properties are not uniform and can depend on the observable under consideration. Our results show that, in photodoped Mott insulators, a second-order strong-coupling expansion may suffice to determine the superconducting transition temperature, whereas a causal description of the Green's functions requires at least third order. The different convergence behavior of thermodynamic phase boundaries and dynamical correlation functions highlights the nontrivial role of higher-order diagrams in nonequilibrium strongly correlated superconductors and underscores the need for controlled 
 impurity solvers when addressing spectroscopic properties.

\begin{figure}[t]{
    \centering
    \includegraphics[width=0.46\textwidth]{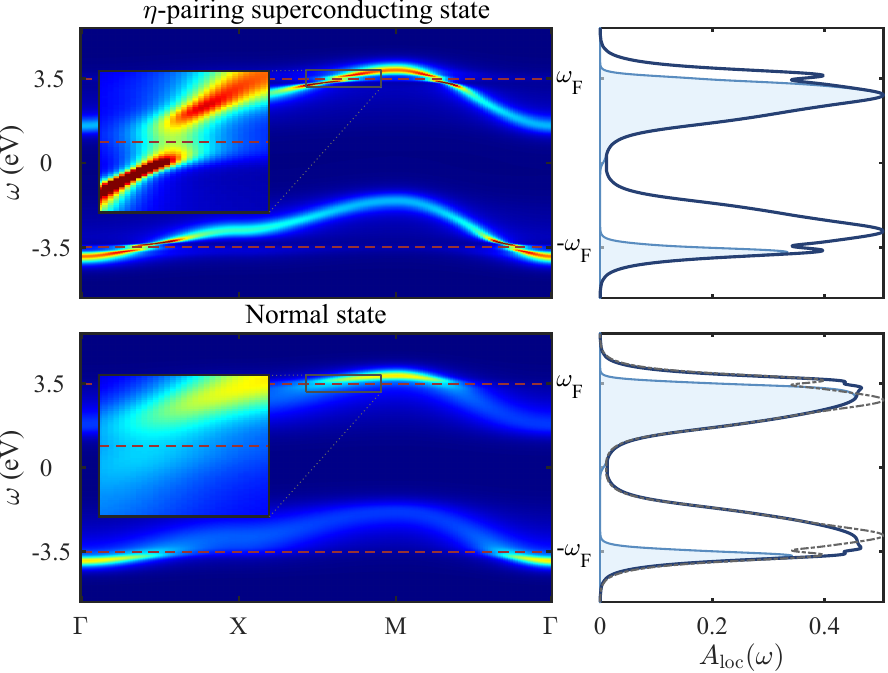}
\caption{Local and momentum-resolved spectra of the photodoped Hubbard model at $T^\text{eff}=1195\,\mathrm{K}$ for a fixed photodoping level corresponding to $\omega_F=3.5\,\mathrm{eV}$ ($\delta \approx 0.4$). 
Top panels: superconducting solution. 
Left: momentum-resolved spectral function along the high-symmetry path $\Gamma$--$X$--$M$--$\Gamma$; the inset shows a magnified view near $\omega_F=3.5\,\mathrm{eV}$ (red dashed line), highlighting the gap opening. 
Right: corresponding local spectral function together with the occupation. 
Bottom panels: results obtained for identical parameters but with the anomalous component suppressed (same color scale).
For comparison, in the right panel, the superconducting local spectrum is overlaid as a gray dash-dotted line.
}
\label{fig_dispersion}
}
\end{figure}

We focus on the photodoped regime at $U=16t$ in the following. To elucidate the spectral signatures of the $\eta$-pairing state, Fig.~\ref{fig_dispersion} shows the local and momentum-resolved spectral functions computed with TOA for a representative photodoped case at $T^\text{eff}=1195\,\mathrm{K}$ and $\omega_F=3.5\,\mathrm{eV}$, corresponding to $\delta\approx0.4$. The top row displays the superconducting solution, while the bottom row shows results obtained for identical parameters, but with the anomalous (pairing) component suppressed. 
In both cases, pronounced quasiparticle peaks appear near the two effective Fermi levels, reflecting the metallic character induced by photodoping. Allowing for pairing, however, leads to clear qualitative changes: the quasiparticle bands become sharper and acquire enhanced spectral weight. As shown in the inset of the top-left panel, pairing further opens a gap at $\omega_F$, resulting in a depletion of spectral weight near the effective Fermi level and a corresponding redistribution in the local spectral function.

Such a superconducting gap has not been reported in previous studies of $\eta$-pairing states in one-dimensional systems~\cite{murakami2022exploring,kaneko2019photoinduced} or on the Bethe lattice~\cite{Li2020}. However, as shown in the SM \cite{SM}, 
similar gap openings are also found at the TOA level in the three-dimensional Hubbard model and for the Bethe lattice with large $\eta$-pairing order parameter. We thus attribute the discrepancies to the accuracy of the solver and to  specificities of the one-dimensional model.  

\begin{figure}[t]{
    \centering
    \includegraphics[width=0.46\textwidth]{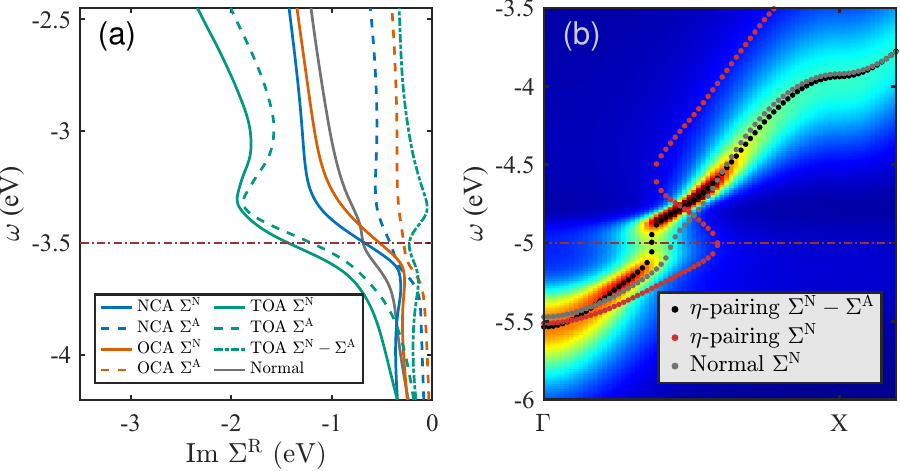}
\caption{(a) Imaginary parts of the retarded normal and anomalous self-energies,
$\Sigma^{\mathrm{N}}(\omega)$ (solid lines) and $\Sigma^{\mathrm{A}}(\omega)$ (dashed lines),
near $\omega=-3.5\,\mathrm{eV}$ for representative $\eta$-pairing superconducting states.
Results obtained with NCA, OCA, and TOA are shown in blue, orange, and green, respectively,
at $T^\text{eff}=1041\,\mathrm{K}$ (NCA, $T_c^{\mathrm{eff}}=1060\,\mathrm{K}$),
$1413\,\mathrm{K}$ (OCA, $T_c^{\mathrm{eff}}=1444\,\mathrm{K}$),
and $1195\,\mathrm{K}$ (TOA, $T_c^{\mathrm{eff}}=1414\,\mathrm{K}$).
For the TOA result, the dash-dotted green line additionally shows $\text{Im}[\Sigma^{\mathrm{N}}(\omega)-\Sigma^{\mathrm{A}}(\omega)$].
All systems are close to the causality-violating regime.
For comparison, the normal self-energy obtained with TOA under identical parameters but with pairing suppressed is shown in gray.
(b) Momentum-resolved spectral function for the $\eta$-pairing state shown in Fig.~\ref{fig_dispersion}.
Dotted lines indicate the quasiparticle band determined from
$\omega-\varepsilon_{\mathbf{k}}-\mathrm{Re}\,\Sigma^{N}(\omega)
+\mathrm{Re}\,\Sigma^{A}(\omega)=0$.
Black dots include $\mathrm{Re}\,\Sigma^{A}(\omega)$,
red dots exclude it,
and gray dots correspond to the normal-state solution. 
}
\label{fig_sig}
}
\end{figure}

The self-energy encodes the many-body effects in the single-particle Green's function and provides a fundamental perspective on spectral properties and causality. In Fig.~\ref{fig_sig}(a), we show the imaginary parts of the retarded normal and anomalous self-energies obtained within TOA for the two cases discussed in Fig.~\ref{fig_dispersion}, together with two additional examples computed using NCA and OCA at the same photodoping level.
All superconducting solutions are chosen close to the regime where the causality constraint~\eqref{causality} is about to be violated for certain frequencies.
A qualitative distinction between the impurity solvers becomes apparent in the vicinity of the effective Fermi level (we focus on $\omega\approx -\omega_F$ in the figure).
In the NCA and OCA results, the normal and anomalous self-energies rapidly approach and intersect already for relatively small order parameters, reflecting their limited capability to consistently describe the $\eta$-pairing state. 
In contrast, the TOA yields a well-separated structure of the normal and anomalous components, which remains stable even far below the effective transition temperature 
$T_c^{\mathrm{eff}}$.

While the imaginary part of the self-energy governs the quasiparticle lifetime, the real part controls the renormalized quasiparticle dispersion.
In the $\eta$-pairing state, the positions of quasiparticle peaks can be estimated from the generalized band equation
$
\omega-\varepsilon_{\mathbf{k}}-\mathrm{Re}\,\Sigma^{N}(\omega)
\pm \mathrm{Re}\,\Sigma^{A}(\omega)=0,
$
as derived in the SM~\cite{SM}.
Using the two examples shown in Fig.~\ref{fig_dispersion}, we extract the corresponding quasiparticle dispersions for both the $\eta$-pairing and normal states. The resulting band positions are overlaid in Fig.~\ref{fig_sig}(b) on top of the superconducting spectral function and are found to coincide remarkably well with the maxima of the spectral weight in the superconducting state. Neglecting the anomalous contribution $\mathrm{Re}\,\Sigma^{A}(\omega)$ leads to substantial deviations, demonstrating the essential role of the anomalous self-energy in shaping the quasiparticle bands.

The broadening of the peaks along these quasiparticle branches is controlled by the imaginary parts of the self-energies, $\mathrm{Im}\,\Sigma^{N}(\omega)\pm\mathrm{Im}\,\Sigma^{A}(\omega)$ (with the minus sign relevant for the branch considered in Fig.~\ref{fig_sig})~\cite{SM}.
This explains the life-time variations observed in Fig.~\ref{fig_sig}(b). 
In particular, close to the gap edges, the $\eta$-pairing state exhibits sharp quasiparticle peaks, despite both the normal and anomalous self-energies being individually large. The difference between the two imaginary components is minimized (see dash-dotted line), leading to a reduced scattering rate. Further away from the gap energy, $\mathrm{Im}\,[\Sigma^{N}(\omega)-\Sigma^{A}(\omega)]$ grows and the bands get blurred. However, this difference remains significantly smaller in the superconducting state than $\mathrm{Im}\,\Sigma^{N}(\omega)$ of the normal-state solution, which explains why in Fig.~\ref{fig_dispersion}, the energy bands of the $\eta$-pairing state are much sharper than those of the normal state. The sharpening of the bands is hence another fingerprint of the transition into the $\eta$-pairing state.

\begin{figure}[t]{
    \centering
    \includegraphics[width=0.46\textwidth]{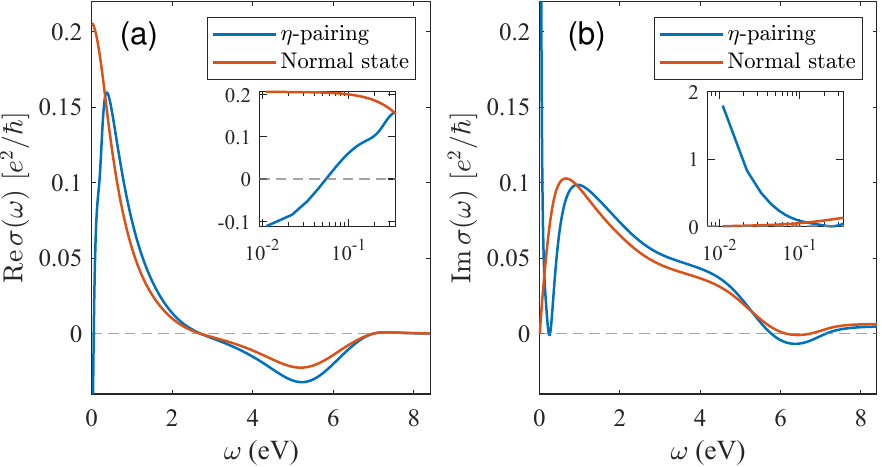}
\caption{Optical conductivity of the photodoped Hubbard model for the same parameters as in Fig.~\ref{fig_dispersion}.
(a) Real part of the optical conductivity, $\mathrm{Re}\,\sigma(\omega)$, excluding the zero-frequency $\delta$-peak contribution.
(b) Imaginary part of the optical conductivity, $\mathrm{Im}\,\sigma(\omega)$.
The low-frequency behavior is shown in the insets on a logarithmic frequency scale.
Blue lines correspond to the $\eta$-pairing superconducting state, while red lines show the normal-state results obtained by suppressing the anomalous component.
}
\label{fig_conductivity}
}
\end{figure}

The optical conductivity provides a particularly sensitive probe of superconductivity. In a normal metallic state, the imaginary part of the optical conductivity vanishes in the zero-frequency limit, whereas in a superconducting state it diverges as $\mathrm{Im}\,\sigma(\omega)\sim 1/\omega$, reflecting the emergence of a finite superfluid stiffness. To provide further experimentally relevant signatures, we show in Fig.~\ref{fig_conductivity} the optical conductivity for the same $\eta$-pairing and normal states as in Fig.~\ref{fig_dispersion}. Details of the calculation are described in the SM~\cite{SM}. 

As shown in Fig.~\ref{fig_conductivity}(a), the real part of the optical conductivity in the $\eta$-pairing state exhibits a pronounced suppression at low but nonzero frequencies, compared to the normal state, signaling the opening of gaps near the effective Fermi levels and the reduction of low-energy dissipative transport. Notably, in the limit $\omega\to 0$, the real part of the conductivity becomes negative, reflecting the nonequilibrium nature of the steady state. 
Analogous negative $\mathrm{Re}\,\sigma(\omega)$ behavior has been measured in driven correlated materials~\cite{PhysRevX.11.011055}. The imaginary part of the conductivity, shown in Fig.~\ref{fig_conductivity}(b), displays a clear $1/\omega$ divergence at low frequencies in the $\eta$-pairing state, consistent with superconducting response. At higher frequencies, we observe negative $\mathrm{Re}\,\sigma(\omega)$ in the range $3\,\mathrm{eV}\lesssim \omega \lesssim 7\,\mathrm{eV}$ in both the normal and superconducting states, indicating energy gain from doublon-holon recombination processes~\cite{Li2020,Werner2019}. Similar results are found for the three-dimensional model \cite{SM}. 
Taken together, the described features provide clear signatures of nonthermal superconductivity in the photodoped Mott system.

Qualitatively similar optical conductivity signatures have been reported in pump-probe experiments on photoexcited correlated materials.
In particular, experiments on cuprates, fullerides and iron-based superconductors~\cite{cavalleri2018photo,rowe2023resonant,isoyama2021light} observed a transient $1/\omega$-like enhancement of $\mathrm{Im}\,\sigma(\omega)$ accompanied by a suppression of $\mathrm{Re}\,\sigma(\omega)$ in photoinduced states.
While those experiments are for metallic systems and rely on pumping protocols that resonantly drive specific phonon modes to transiently reshape electronic coherence, here charge carriers are photodoped into the Hubbard bands of a Mott insulator and the resulting $\eta$-pairing superconducting state emerges purely from electronic correlations. 
Whether or not long-range $\eta$-pairing correlations can play a role in the short-time evolution of light-driven strongly correlated metals is an interesting topic for future investigations. 

In summary, our work establishes large-gap Mott insulators as a promising platform for light-controlled high-temperature superconductivity. We showed that $T^\text{eff}_c$ can substantially exceed room temperature even at moderate photodoping, suggesting that metastable superconducting states with sufficiently low effective electronic temperatures may be realized through energy dissipation to the lattice or entropy-cooling protocols~\cite{Werner2019,Werner2019light,nava2018cooling}. Clear spectroscopic signatures of the superconducting state were identified in both the momentum-resolved spectral function and the optical conductivity, providing experimentally accessible evidence of $\eta$ pairing. From a methodological perspective, our results demonstrate that a reliable description of the $\eta$-pairing state requires higher-order strong-coupling impurity solvers, highlighting the importance of controlled 
approaches for nonequilibrium strongly correlated systems. 

The emergence of robust $\eta$-pairing superconductivity in photodoped Mott insulators opens a route toward light-induced superconductivity that is fundamentally distinct from equilibrium pairing mechanisms. Our findings provide quantitative guidance and concrete benchmarks for ongoing experimental efforts to realize and characterize photoinduced superconducting states in correlated materials.

{\it Acknowledgments}
This work was supported by Swiss National Science Foundation via NCCR Marvel and Grant No.~2000-1-240023, and the DGIST Start-up Fund Program of the Ministry of Science and ICT (2025010006) and Basic Science Research Program through the National Research Foundation of Korea (NRF) funded by the Ministry of Education (2025090055).
The calculations were run on the Beo06 cluster at the University of Fribourg.
\nocite{ginzburg2009theory,Nayak2025,Martin2008,Haule_2007,Ferrell1958,Tinkham1956}
\bibliography{mybibtex}

\end{document}


\title{Supplemental Material for \texorpdfstring{\\}{} High-temperature \texorpdfstring{$\eta$}{eta}-pairing superconductivity in the photodoped Hubbard model}

\author{Lei Geng}
\affiliation{Department of Physics, University of Fribourg, Fribourg-1700, Switzerland}
\author{Aaram J. Kim}
\affiliation{Department of Physics and Chemistry, DGIST, Daegu 42988, Korea}
\author{Philipp Werner}
\affiliation{Department of Physics, University of Fribourg, Fribourg-1700, Switzerland}

\date{\today}
\maketitle

\section{Nambu basis}

The Nambu basis provides a convenient framework for describing superconducting systems 
which break the $U(1)$ symmetry. It treats particle and hole operators on equal footing by combining them into a two-component spinor.  
On a given site, the Nambu spinor is defined as 
\begin{equation}
    \Psi =
    \begin{pmatrix}
        c_{\uparrow} \\
        c^{\dagger}_{\downarrow}
    \end{pmatrix}.
\end{equation}
Using this spinor, the Green's function becomes a $2\times2$ matrix~\cite{Georges1996},
\begin{equation}
    G(t,0) = -\mathrm{i}\,
    \langle \mathcal{T} \Psi(t)\Psi^{\dagger}(0)\rangle =
    \begin{pmatrix}
        G_{\uparrow}(t,0) & F(t,0) \\
        F^*(t,0) & -G_{\downarrow}(0,t)
    \end{pmatrix},
\end{equation}
where $\mathcal{T}$ denotes the time-ordering operator and $G$ and $F$ represent the normal and anomalous components, respectively.  
In practice, the most relevant object is the retarded Green's function; the lesser and greater components describing the electron and hole occupations 
can be obtained from it by taking into account the (externally imposed) distribution function of the steady state. Below, we briefly analyze the symmetry relations among the four matrix elements.

For a steady state, the matrix elements of the retarded Green's function in the time domain read
\begin{equation}
    \begin{aligned}
        G^{\mathrm{R}}_{11}(t) &= -\mathrm{i}\theta(t)\langle c_{\uparrow}(t)c^{\dagger}_{\uparrow}(0)
        + c^{\dagger}_{\uparrow}(0)c_{\uparrow}(t) \rangle,\\
        G^{\mathrm{R}}_{22}(t) &= -\mathrm{i}\theta(t)\langle c^{\dagger}_{\downarrow}(t)c_{\downarrow}(0)
        + c_{\downarrow}(0)c^{\dagger}_{\downarrow}(t) \rangle,\\
        G^{\mathrm{R}}_{12}(t) &= -\mathrm{i}\theta(t)\langle c_{\uparrow}(t)c_{\downarrow}(0)
        + c_{\downarrow}(0)c_{\uparrow}(t) \rangle,\\
        G^{\mathrm{R}}_{21}(t) &= -\mathrm{i}\theta(t)\langle c^{\dagger}_{\downarrow}(t)c^{\dagger}_{\uparrow}(0)
        + c^{\dagger}_{\uparrow}(0)c^{\dagger}_{\downarrow}(t) \rangle .
    \end{aligned}
\end{equation}
For a spin-unpolarized spin-singlet system, the following relations hold:
\begin{equation}
    \begin{aligned}
        G^{\mathrm{R}}_{22}(t) &= -\left[G^{\mathrm{R}}_{11}(t)\right]^*,\\
        G^{\mathrm{R}}_{21}(t) &= \left[G^{\mathrm{R}}_{12}(t)\right]^*.
    \end{aligned}
\end{equation}
After Fourier transformation, we obtain the corresponding frequency-domain relations: 
\begin{equation}
    \begin{aligned}
        G^{\mathrm{R}}_{22}(\omega) &= -\left[G^{\mathrm{R}}_{11}(-\omega)\right]^*, \\
	G^{\mathrm{R}}_{21}(\omega) &= \left[G^{\mathrm{R}}_{12}(-\omega)\right]^*.
    \end{aligned}
\end{equation}
In the special case of particle-hole symmetry and a real order parameter (enforced by a real symmetry-breaking seed), one further has
$G^{\mathrm{R}}_{22}(\omega)=G^{\mathrm{R}}_{11}(\omega)$ and $G^{\mathrm{R}}_{12}(\omega)=G^{\mathrm{R}}_{21}(\omega)$.

For a translationally invariant lattice model, the Nambu basis can be defined analogously in momentum space. Its precise form depends on the symmetry of the ordered state under consideration. For a conventional Cooper pairing state, involving electrons with momentum $\mathbf{k}$ and $-\mathbf{k}$,
the Nambu spinor is defined as
\begin{equation}
    \Psi_{\mathbf{k}} =
    \begin{pmatrix}
        c_{\mathbf{k},\uparrow} \\
        c^{\dagger}_{-\mathbf{k},\downarrow}
    \end{pmatrix}.
\end{equation}
In this basis, the noninteracting Hamiltonian in momentum space can be written as
\begin{equation}
    H_0 =
    \sum_{\mathbf{k}}
    \Psi^{\dagger}_{\mathbf{k}}
    \begin{pmatrix}
        \varepsilon_{\mathbf{k}} & 0 \\
        0 & -\varepsilon_{-\mathbf{k}}
    \end{pmatrix}
    \Psi_{\mathbf{k}},
\end{equation}
where $\varepsilon_{\mathbf{k}}=-2t\left[\cos(k_x)+\cos(k_y)\right]$ for the two-dimensional square lattice. In the main text, this representation is used for the attractive Hubbard model.

In the $\eta$-pairing state on the two-dimensional square lattice, the Cooper pairs carry a finite center-of-mass momentum $\mathbf{Q}=(\pi,\pi)$. Accordingly, the Nambu spinor is chosen as
\begin{equation}
    \Psi_{\mathbf{k}} =
    \begin{pmatrix}
        c_{\mathbf{k},\uparrow} \\
        c^{\dagger}_{\mathbf{Q}-\mathbf{k},\downarrow}
    \end{pmatrix}.
    \label{eq_nambu}
\end{equation}
Using this basis, the noninteracting Hamiltonian in momentum space becomes
\begin{equation}
    H_0 =
    \sum_{\mathbf{k}}
    \Psi^{\dagger}_{\mathbf{k}}
    \begin{pmatrix}
        \varepsilon_{\mathbf{k}} & 0 \\
        0 & -\varepsilon_{\mathbf{Q}-\mathbf{k}}
    \end{pmatrix}
    \Psi_{\mathbf{k}}.
\end{equation}

Within this Nambu representation, a nonzero off-diagonal expectation value of the Green's function corresponds to anomalous pairing correlations between fermions with momenta $\mathbf{k}$ and $\mathbf{Q}-\mathbf{k}$,
\begin{equation}
    F_{\mathbf{k}} =
    \left\langle
        c_{\mathbf{Q}-\mathbf{k},\downarrow}
        c_{\mathbf{k},\uparrow}
    \right\rangle \neq 0,
\end{equation}
which signals the formation of an $\eta$-pairing superconducting state with a finite center-of-mass momentum $\mathbf{Q}$. The physical origin of this pairing structure becomes more transparent when it is examined in real space.

As discussed in previous works~\cite{murakami2025photoinduced,murakami2022exploring}, $\eta$-pairing is characterized by a staggered on-site pairing amplitude in real space,
\begin{equation}
    \left\langle c_{i\downarrow} c_{i\uparrow} \right\rangle
    \propto e^{i\mathbf{Q}\cdot\mathbf{R}_i},
    \qquad
    \mathbf{Q}=(\pi,\pi),
\end{equation}
which describes local Cooper pairing with an alternating sign between neighboring lattice sites. Introducing the Fourier transform of the fermionic operators,
\begin{equation}
    c_{i,\sigma}
    =
    \frac{1}{\sqrt{N_k}}
    \sum_{\mathbf{k}}
    e^{i\mathbf{k}\cdot\mathbf{R}_i}
    c_{\mathbf{k},\sigma},
\end{equation}
the anomalous expectation value in real space can be expressed as
\begin{align}
    \left\langle c_{i\downarrow} c_{i\uparrow} \right\rangle
    &=
    \frac{1}{N_k}
    \sum_{\mathbf{k}',\mathbf{k}}
    e^{i(\mathbf{k}'+\mathbf{k})\cdot\mathbf{R}_i}
    \left\langle
        c_{\mathbf{k}'\downarrow}
        c_{\mathbf{k}\uparrow}
    \right\rangle .
\end{align}
Comparison with the staggered form proportional to $e^{i\mathbf{Q}\cdot\mathbf{R}_i}$ shows that the anomalous expectation value is restricted to the channel satisfying
\begin{equation}
    \mathbf{k}' + \mathbf{k} = \mathbf{Q},
\end{equation}
which leads to
\begin{equation}
    \left\langle
        c_{\mathbf{Q}-\mathbf{k},\downarrow}
        c_{\mathbf{k},\uparrow}
    \right\rangle \neq 0.
\end{equation}
This observation justifies the choice of the Nambu spinor involving momenta $\mathbf{k}$ and $\mathbf{Q}-\mathbf{k}$.

It is important to emphasize that the restriction to pairing between momenta $\mathbf{k}$ and $\mathbf{Q}-\mathbf{k}$ in momentum space does not imply that pairing correlations are purely local in real space. In particular, the anomalous expectation value between different lattice sites, $\langle c_{i,\downarrow} c_{j,\uparrow} \rangle$ with $i\neq j$, is in general nonzero. A more general anomalous correlator can be written as
\begin{align}
    \left\langle c_{i,\downarrow} c_{j,\uparrow} \right\rangle
    &=
    \frac{1}{N_k}
    \sum_{\mathbf{k}}
    e^{i\mathbf{k}\cdot(\mathbf{R}_i-\mathbf{R}_j)}
    e^{i\mathbf{Q}\cdot \mathbf{R}_j}
    F_{\mathbf{k}} \nonumber\\
    &=
    e^{i\mathbf{Q}\cdot \mathbf{R}_j}
    \tilde{F}(\mathbf{R}_i-\mathbf{R}_j),
\end{align}
where $\tilde{F}(\mathbf{R})$ encodes the dependence of the anomalous correlator on the relative coordinate. Such spatially extended pairing correlations naturally arise from a purely local staggered order parameter once band dispersion and fermionic propagation are taken into account. This holds even though the self-energy remains local within the DMFT framework.


This observation also guides our choice of formulation. Since the pairing correlations extend beyond the on-site component, a representation based on a single-site Nambu spinor with two components (Eq.~\eqref{eq_nambu})
provides the most consistent way to incorporate superconductivity while preserving the locality of the self-energy assumed in DMFT. If one instead adopts a two-sublattice ($A/B$) formulation with a four-component spinor analogous to antiferromagnetic DMFT~\cite{Geng2025third}, inter-sublattice anomalous contributions cannot be treated on equal footing with local correlations and are suppressed, which introduces additional approximations.  Although such a formulation may produce similar transition temperatures than the formalism based on Eq.~\eqref{eq_nambu}, it leads to noticeable changes in the spectral functions.

\section{DMFT loop}
\label{dmftloop}

The steady-state DMFT calculations presented in the main text employ the high-order strong-coupling impurity solvers developed in Ref.~\cite{Geng2025third}. Accordingly, the impurity part of the DMFT self-consistency loop follows the scheme introduced there. However, in Ref.~\cite{Geng2025third}, the calculations were performed on the infinite-dimensional Bethe lattice, for which the hybridization function can be obtained directly from the local Green's function, without extracting the self-energy of the impurity model.   

In contrast, for the two- and three-dimensional lattices considered in the present work, the local lattice Green's function, which determines the hybridization function, must be computed through an explicit momentum integration. This requires several additional steps in the self-consistency procedure. Below, we describe the complete DMFT loop used in this study.

\begin{enumerate}
\item Starting from an initial guess (in the first iteration) or from the solution of the previous iteration for the pseudo-particle Green's functions and the hybridization function, the pseudo-particle self-energies are evaluated diagrammatically, including both the normal and anomalous contributions.
  
\item The steady-state Dyson equations for the pseudo-particles are solved in frequency space to obtain the pseudo-particle Green's functions from the corresponding self-energies~\cite{kim2025strong,li2021nonequilibrium}.
  
\item The local physical Green's function $G_{\mathrm{loc}}$, with normal and anomalous components, is calculated from the updated pseudo-particle Green's functions and the hybridization function according to the diagrammatic rules of the strong-coupling expansion. The lesser and greater components are obtained by imposing an effective temperature distribution for the doublons (holons) in the upper (lower) Hubbard bands~\cite{kunzel2024numerically}.
  
\item The hybridization function is updated using the newly obtained local lattice Green's function and the hybridization function from the previous iteration, following the steps below:
\begin{enumerate}[label=\alph*.]
    \item The ``atomic'' Green's function $G_1$ is constructed from the local physical Green's function $G_{\mathrm{loc}}$ and the old hybridization function $\Delta$ via
    \begin{equation}
    G_1^{-1} = G_{\mathrm{loc}}^{-1} + \Delta .
    \end{equation}
    \item The momentum-dependent Green's function $G_{\mathbf{k}}$ is obtained from $G_1$ and the noninteracting Hamiltonian in momentum space: 
    \begin{equation}
    G_{\mathbf{k}}^{-1} = G_1^{-1} - H_0(\mathbf{k}) .
    \end{equation}
    Note that terms related to the chemical potential are incorporated into $G_1$ rather than $H_0(\mathbf{k})$.
    \item The updated local Green's function $G_2$ is computed by averaging over momentum in the first Brillouin zone,
    \begin{equation}
    G_2 = \frac{1}{N_k} \sum_{\mathbf{k}} G_{\mathbf{k}} .
    \end{equation}
    \item The new hybridization function is determined as
    \begin{equation}
    \Delta_{\mathrm{new}} = G_1^{-1} - G_2^{-1} .
    \end{equation}
\end{enumerate}
\end{enumerate}

The above procedure is iterated until convergence of the Green's functions and the hybridization function is achieved.

\section{Determination of the superconducting phase boundary}

To determine the superconducting transition temperature, we employ a procedure which combines a seed field with extrapolation. 
Specifically, we add a small anomalous pairing field to the impurity Hamiltonian,
\begin{equation}
H_{\mathrm{seed}} = -h_{\mathrm{pair}}
\left(c_{\downarrow}c_{\uparrow} + \mathrm{H.c.}\right),
\end{equation}
with $h_{\mathrm{pair}} = 10^{-4}$. This seed explicitly breaks the $U(1)$ symmetry and stabilizes the superconducting solution, 
while being sufficiently small so as not to affect the order parameter far enough from the critical point. In fact, for the solutions considered, the superconducting state remains robust even if the seed field is withdrawn after initial convergence, with the order parameter showing only a negligible change.

\begin{figure}[t]
\includegraphics[width=0.45\textwidth]{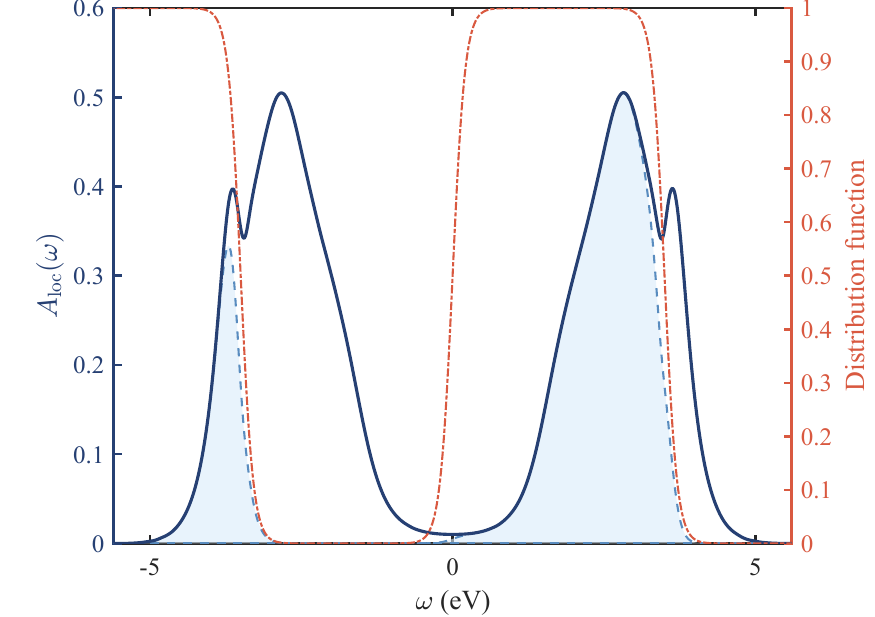}
\caption{
Distribution function for the $\eta$-pairing state (orange dash-dotted line). The spectral function and occupation from Fig.~2 of the main text are reproduced for reference.
}
\label{fig_distribution}
\end{figure}

The effective temperature $T^\text{eff}$ determines the Fermi--Dirac-like distributions imposed in our steady-state setup. Figure~\ref{fig_distribution} shows the distribution function for the $\eta$-pairing state shown in Fig.~2 of the main text. In the energy range of the upper and lower Hubbard bands, the distribution is well described by Fermi--Dirac functions centered at $\pm \omega_F$, with temperature $T^\text{eff}$,  $f(\omega)\approx 1/(\exp[(\omega\mp\omega_F)/T^\text{eff}]+1)$.  Around $\omega=0$, a smooth crossover connects these two Fermi--Dirac functions. However, this crossover does not influence the physics discussed here, as it lies within the Mott gap~\cite{Geng2025third}.
For several such effective temperatures $T^\text{eff}$, we solve the DMFT equations self-consistently and measure the anomalous expectation value
\(
\Delta \equiv \langle c_{\downarrow}c_{\uparrow}\rangle
\),
which serves as the superconducting order parameter. Close to the transition, the order parameter follows the mean-field critical behavior~\cite{ginzburg2009theory}
\begin{equation}
\Delta(T^\text{eff}) \propto \sqrt{T^\text{eff}_c - T^\text{eff}}.
\end{equation}
We therefore determine $T^\text{eff}_c$ by fitting the temperature-dependent data to this square-root form and extrapolating $\Delta(T^\text{eff})$ to zero.

\begin{figure}[t]
\includegraphics[width=0.45\textwidth]{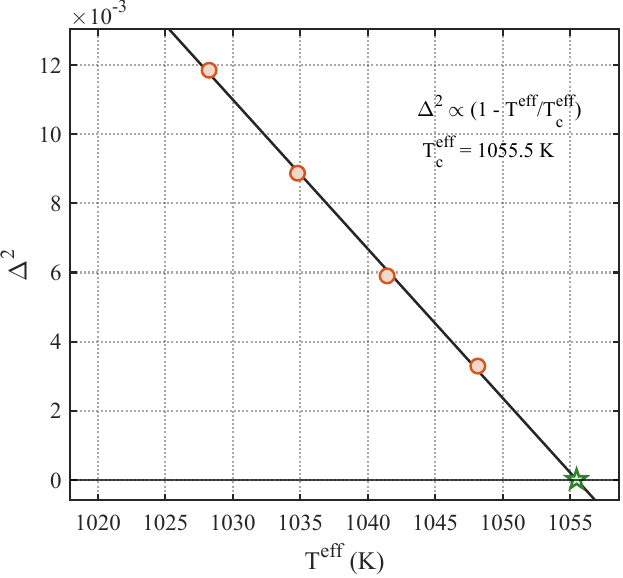}
\caption{
Extrapolation procedure for determining the superconducting transition temperature $T^\text{eff}_c$ of the photodoped Hubbard model. The squared order parameter $\Delta^2$ (red circles) is plotted as a function of temperature $T^\text{eff}$ at a fixed photodoping level $\omega_F = 3.78\,\mathrm{eV}$, corresponding to an effective doping concentration of $\delta \approx 0.43$. The data shown are calculated for the square lattice using the NCA impurity solver. The solid black line represents a linear fit consistent with the Ginzburg-Landau mean-field behavior $\Delta^2 \propto (T^\text{eff}_c - T^\text{eff})$. Extrapolating the fit to zero (green star) yields a transition temperature of $T^\text{eff}_c = 1055.5$\,K.
}
\label{fig_fitting}
\end{figure}

In practice, we perform the fit using several temperature points in the regime where $\Delta$ is small. As illustrated in Fig.~\ref{fig_fitting}, the square-root scaling provides an excellent description of the data, which allows for an accurate and robust extraction of $T_c$. We verified that further reducing the seed amplitude or varying the fitting window does not significantly change the resulting transition temperature.

\section{Causality constraints and limitations of low-order impurity solvers for the \texorpdfstring{$\eta$}{eta}-pairing state}

As discussed in our previous work~\cite{Geng2025third}, in generic Mott insulating regimes the OCA already provides reasonably accurate spectral functions, while the TOA yields only quantitative improvements. However, the situation changes qualitatively in the photodoped Mott regime considered here. The electronic state simultaneously hosts strong local correlations inherited from the Mott background and itinerant carriers generated by photoexcitation. This coexistence places high requirements on the accuracy and numerical stability of impurity solvers within DMFT. 

In this regime, we found that the TOA spectral functions differ substantially from those obtained with NCA and OCA, while remaining in close agreement with the recently reported results from the influence functional (IF) method~\cite{Geng2025third,Nayak2025}. This indicates that higher-order crossing diagrams become essential when strong correlations coexist with photoinduced carriers, and that lower-order strong-coupling expansions are no longer sufficient.

The limitations of NCA and OCA become even more severe for the $\eta$-pairing states studied in this work. In the superconducting phase, there are additional constraints on the anomalous components of the Green's function. If these constraints are not satisfied with sufficient accuracy, the DMFT loop becomes unstable and fails to converge. In practice, both NCA and OCA encounter convergence problems already at temperatures slightly below the $\eta$-pairing transition temperature. Although phase boundaries can still be inferred from the order parameter, these solvers cannot access the low-temperature regime where characteristic spectral features of the superconducting state emerge. Consequently, reliable spectroscopic information in the ordered phase cannot be obtained within these approximations.

By contrast, the TOA remains numerically stable down to substantially lower temperatures and yields fully converged solutions deep inside the superconducting phase. This robustness enables a characterization of the spectral properties of the $\eta$-pairing state, including the opening of the superconducting gap and the associated redistribution of spectral weight shown in the main text. We therefore conclude that at least third-order strong-coupling corrections, as implemented in the TOA, are necessary for a quantitatively reliable description of photodoped Mott systems and, in particular, for resolving the spectroscopic fingerprints of $\eta$-pairing superconductivity.

In the following, we briefly summarize the additional physical constraints that arise in the superconducting phase. As discussed in Ref.~\cite{Yue2023}, for a broad class of interacting systems one can construct an auxiliary self-energy whose spectral function is positive semidefinite, using suitable linear combinations of the normal and anomalous self-energies. For the particle-hole symmetric system considered here, this general result simplifies. Denoting the normal and anomalous components of the self-energy in Nambu space by $\Sigma^{\mathrm{N}}(\omega)$ and $\Sigma^{\mathrm{A}}(\omega)$, respectively, one obtains
\begin{equation}
\left| \mathrm{Im}\,\Sigma^{\mathrm{N}}(\omega) \right|
\;\ge\;
\left| \mathrm{Im}\,\Sigma^{\mathrm{A}}(\omega) \right|,
\qquad \forall\,\omega ,
\label{eq:causality_constraint}
\end{equation}
i.e., the magnitude of the imaginary part of the normal self-energy must dominate over that of the anomalous component at all frequencies. In addition, the normal constraint $\mathrm{Im}\,\Sigma^{\mathrm{N}}(\omega)<0$ still applies.

This constraint is related to the causality of the solution. 
In Nambu space, the retarded Green's function satisfies the Dyson equation
\begin{align}
	G^{-1}_\mathbf{k}(\omega)
	&= G_{0,\mathbf{k}}^{-1}(\omega)-\Sigma(\omega) \nonumber\\
	&=
\begin{pmatrix}
	\omega - \varepsilon_\mathbf{k} - \Sigma^{\mathrm{N}}(\omega) & -\Sigma^{\mathrm{A}}(\omega) \\
	-\Sigma^{\mathrm{A}}(\omega) & \omega + \varepsilon_\mathbf{{Q-k}} + {\Sigma^{\mathrm{N}*}(-\omega)}
\end{pmatrix}.
\label{dyson_equation}
\end{align}
After applying the particle-hole symmetry, 
\begin{equation}
	G^{-1}_\mathbf{k}(\omega) = 
	\left(
	\begin{array}{cc}
		\omega - \varepsilon_\mathbf{k} - \Sigma^{\mathrm{N}}(\omega) & -\Sigma^{\mathrm{A}}(\omega) \\
		-\Sigma^{\mathrm{A}}(\omega) & \omega - \varepsilon_\mathbf{{k}} - {\Sigma^{\mathrm{N}}(\omega)}
	\end{array}
\right)~.
	\label{<+label+>}
\end{equation}
Causality requires the spectral function 
$A_\mathbf{k}(\omega)=-\mathrm{Im}\,G_\mathbf{k}(\omega)/\pi$ 
to be a positive semidefinite matrix. 
Using the identity
\begin{equation}
\begin{aligned}
G_\mathbf{k}^{-1}-\left(G_\mathbf{k}^{-1}\right)^{\dagger}
&=
-\,G_\mathbf{k}^{-1}\left(G_\mathbf{k}-G_\mathbf{k}^{\dagger}\right)\left(G_\mathbf{k}^{-1}\right)^{\dagger},\\
2\,\mathrm{Im}\,G_\mathbf{k}^{-1}
&=
-2\,G_\mathbf{k}^{-1}\left(\mathrm{Im}\,G_\mathbf{k}\right)\left(G_\mathbf{k}^{-1}\right)^{\dagger},
\end{aligned}
\end{equation}
one sees that if $-\mathrm{Im}\,G_\mathbf{k}(\omega)$ is positive semidefinite, then $\mathrm{Im}\,G_\mathbf{k}^{-1}(\omega)$ must also be positive semidefinite. 
Since $\mathrm{Im}\,G_{0,\mathbf{k}}^{-1}=0$, the Dyson equation implies that $-\mathrm{Im}\,\Sigma(\omega)$ is likewise positive semidefinite. 
For the $2\times2$ Nambu matrix in Eq.~\eqref{dyson_equation}, this condition is equivalent to
\begin{equation}
\bigl(\mathrm{Im}\,\Sigma^{\mathrm{N}}(\omega)\bigr)^2
-
\bigl(\mathrm{Im}\,\Sigma^{\mathrm{A}}(\omega)\bigr)^2
\ge 0,\qquad \mathrm{Im}\,\Sigma^{\mathrm{N}}(\omega)<0,
\end{equation}
and the first condition directly yields Eq.~\eqref{eq:causality_constraint}. 
Therefore, the inequality above constitutes a necessary and sufficient condition for the Green's function to remain causal.

In practice, we find that NCA and OCA violate this constraint already at temperatures slightly below the superconducting transition temperature. This violation leads to noncausal spectral functions and, consequently, unstable DMFT iterations. By contrast, the TOA respects Eq.~\eqref{eq:causality_constraint} over the entire frequency range down to substantially lower temperatures, which explains its superior numerical stability and reliability in the superconducting phase.

\begin{figure}[t]
\includegraphics[width=0.45\textwidth]{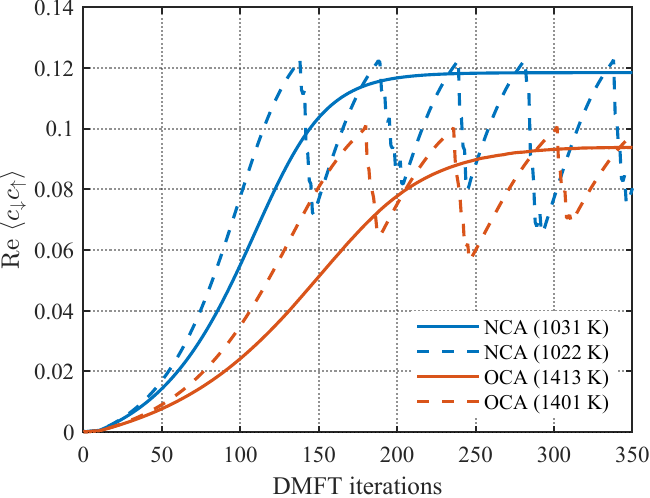}
\caption{
Evolution of the superconducting order parameter 
$\mathrm{Re}\,\langle c_{\downarrow} c_{\uparrow} \rangle$ 
as a function of the DMFT iteration number for photodoped 
$\eta$-pairing states at doping $\delta \approx 0.4$, obtained with the NCA and OCA impurity solvers for the square lattice. 
Blue curves show NCA results (solid: $T^\text{eff}=1031\,\mathrm{K}$, dashed: $1022\,\mathrm{K}$), while red curves show OCA results (solid: $1413\,\mathrm{K}$, dashed: $1401\,\mathrm{K}$). 
All cases are chosen close to the causality constraint~\eqref{eq:causality_constraint}. Solid lines correspond to temperatures where the constraint is still satisfied and the DMFT loop converges to a stable solution, whereas dashed lines indicate slight violations of the constraint, leading to nonconvergent iterations and the breakdown of the superconducting solution.
}
\label{fig_collapse}
\end{figure}

As an illustration, Fig.~\ref{fig_collapse} shows the evolution of the expectation value $\mathrm{Re}\,\langle c_{\downarrow}c_{\uparrow}\rangle$ during the DMFT iterations for systems close to the violation of the causality constraint~\eqref{eq:causality_constraint}, at a photodoping level $\delta \approx 0.4$. For each method (NCA and OCA), two temperatures are considered. In the higher-temperature case, which respects the causality constraint, the iterations converge smoothly. In contrast, at slightly lower temperatures the iterations exhibit repeated collapses and recoveries, signaling an imminent breakdown of causality. Inspection of the self-energies near this instability reveals that the anomalous and normal components become comparable in magnitude, indicating that the causality constraint is about to be violated. Notably, OCA even loses stability at a smaller value of the order parameter than NCA, demonstrating that the breakdown point is not a monotonic function of the diagrammatic order. For reference, the phase transition temperatures at this photodoping are $T^\text{eff}_c = 1065$~K (NCA), 1444~K (OCA), and 1414~K (TOA). Both NCA and OCA fail to converge roughly 40~K below their respective transition temperatures, whereas TOA remains stable down to 1160~K—more than 250~K below its transition temperature. This marked extension of the stable superconducting regime shows that the step from OCA to TOA is qualitatively important for the description of $\eta$-pairing states with large order parameter.

\begin{figure}[t]
\includegraphics[width=0.45\textwidth]{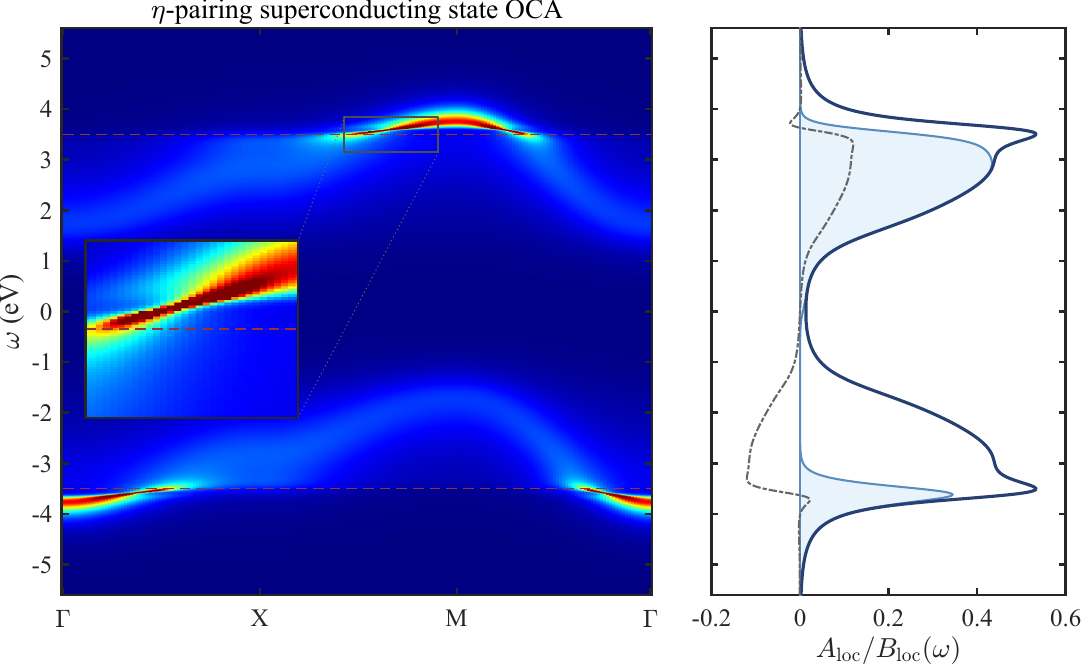}
\caption{
Momentum- and frequency-resolved normal spectra for the $\eta$-pairing state obtained using the OCA at $T^\text{eff}=1413\,\mathrm{K}$ and a photodoping level $\delta \approx 0.4$ ($\omega_F= 3.5\,\mathrm{eV}$).
Left panel: momentum-resolved normal spectral function for the square lattice along the high-symmetry path $\Gamma$--$X$--$M$--$\Gamma$.
Right panel: local normal spectral function $A_{\mathrm{loc}}(\omega)$ together with the corresponding occupation.
For comparison, the anomalous local spectral function is overlaid as a gray dash-dotted line.
}
\label{fig_OCA}
\end{figure}

To further assess the behavior of low-order impurity solvers for $\eta$-pairing states, we analyze the spectral functions obtained from the OCA solution.
The results are shown in Fig.~\ref{fig_OCA}, with the momentum-resolved spectra displayed in the left panel and the corresponding local spectral function in the right panel.
Strikingly, even in the parameter regime close to the causality-violating limit, neither the momentum-resolved spectra nor the local spectral function exhibit any signatures of a superconducting gap.
Instead, a maximum of the spectral weight appears above the effective Fermi energy $\omega_F$, corresponding to the touching point of the normal and anomalous self-energies shown in Fig.~3 of the main text. 
This behavior highlights a fundamental limitation of low-order strong-coupling impurity solvers such as NCA and OCA.
Their breakdown occurs before physically meaningful superconducting features, such as a gap opening, can be established in the single-particle spectra.
For the purpose of calculating the spectra, it is hence important to employ the TOA solver, which gives access to the physically relevant deep $\eta$-pairing regime.

Finally, we comment on a subtle but important difference that arises for calculations performed on the Bethe lattice. As discussed above, when the DMFT loop is implemented following the standard lattice procedure involving the self-energy, a violation of the causality constraint~\eqref{eq:causality_constraint} inevitably produces noncausal spectral functions with negative weight, and the self-consistency cycle consequently becomes unstable and fails to converge. In Bethe-lattice calculations, however, this issue is not immediately apparent.

The origin of this behavior can be traced to the special form of the DMFT self-consistency on the Bethe lattice. 
For a semi-elliptical density of states, the lattice Dyson equation can be evaluated analytically, and the momentum summation reduces to the closed relation~\cite{Georges1996}
\begin{equation}
\Delta(\omega)=t^{2} G_{\mathrm{loc}}(\omega),
\end{equation}
which expresses the hybridization function directly in terms of the local Green's function and the (renormalized) hopping parameter $t$. Consequently, the self-consistency loop updates $\Delta$ without explicitly reconstructing the momentum-resolved lattice Green's functions, thereby bypassing the usual lattice step of the DMFT procedure. 

This shortcut relies on the Hilbert transform of the semi-elliptical density of states and is mathematically justified only when the local Green's function is itself causal. Because the momentum-dependent Green's functions are never constructed, violations of the self-energy causality constraint are not directly exposed, and the iteration may converge to an unphysical fixed point whose spectral function may look reasonable, but does not correspond to any consistent lattice Hamiltonian.

By explicitly comparing the two implementations on the Bethe lattice, i.e., the simplified update $\Delta(\omega)=t^{2} G_{\mathrm{loc}}(\omega)$ and the explicit DMFT procedure involving the calculation of the self-energy and integration over the semi-elliptical density of states, 
we found that both approaches yield indistinguishable results when Eq.~\eqref{eq:causality_constraint} is satisfied. Upon lowering the temperature, however, qualitative differences emerge: the explicit scheme ceases to converge, indicating the breakdown of causality, whereas the simplified $\Delta=t^{2}G$ iteration continues to produce apparently well-behaved solutions. This behavior exemplifies the ``false convergence'' discussed above, where numerical stability masks an underlying violation of physical constraints. Bethe lattice solutions should therefore be interpreted with particular caution.

\section{Momentum-resolved occupation and anomalous spectra}

In the main text, we present momentum- and frequency-resolved spectral functions obtained from the imaginary part of the retarded Green's function, together with the corresponding local spectra and electronic occupations.
While this representation provides direct access to the excitation spectrum and the superconducting gap, angle-resolved photoemission spectroscopy (ARPES) experiments probe only the momentum-resolved occupied part of the spectrum, i.e., $- \frac{1}{\pi}\mathrm{Im}\, G^{<}(\mathbf{k},\omega) = f(\omega)\, A(\mathbf{k},\omega)$, where $f(\omega)$ is the distribution function.

\begin{figure}[t]{
    \centering
    \includegraphics[width=0.46\textwidth]{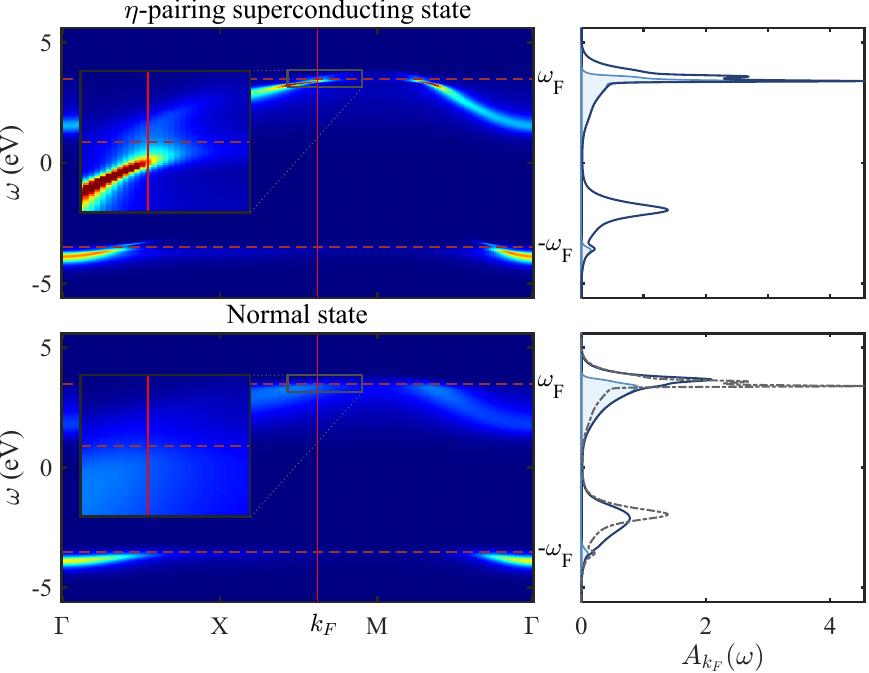}
\caption{Momentum- and frequency-resolved occupation of the photodoped Hubbard model at $T^\text{eff}=1195\,\mathrm{K}$ for the same parameters as in Fig.~2 of the main text.
Left panels: momentum-resolved occupied spectral function for the square lattice, $- \frac{1}{\pi}\mathrm{Im}\, G^{<}(\mathbf{k},\omega)$, along the high-symmetry path $\Gamma$--$X$--$M$--$\Gamma$.
The red solid line marks the representative momentum $k_F$ close to the gap-opening region.
Right panels: spectral function and corresponding occupation evaluated at the selected momentum $k_F$.
Top panels show the superconducting solution, while bottom panels correspond to the normal-state solution obtained by suppressing the anomalous component.
For comparison, the superconducting spectra are overlaid as gray dash-dotted lines in the bottom-right panel.
}
\label{fig_dispersion_les}
}
\end{figure}

To facilitate a direct comparison with ARPES measurements, we therefore show in Fig.~\ref{fig_dispersion_les} the momentum- and frequency-resolved occupied spectra computed for the same parameters as Fig.~2 of the main text. As expected, the superconducting gap identified in the spectral function is not directly visible in the occupied spectra, since only energies below the Fermi level contribute.
Nevertheless, the formation of the superconducting state manifests itself through characteristic changes in the spectral shape.

In particular, upon entering the superconducting phase, the quasiparticle weight in the occupied states slightly below the Fermi energy is significantly enhanced, leading to the emergence of pronounced coherence peaks. This enhancement reflects a longer quasiparticle lifetime and an increased  electronic coherence induced by pairing. Such effects are especially obvious near the region where the gap opens in the momentum-resolved spectra. To highlight this behavior, we select a representative momentum $k_F$ located in this regime, as indicated by the red solid line in the left panels of Fig.~\ref{fig_dispersion_les}. The spectral function and occupation at this momentum are shown in the right panels. Compared to the local spectra in the main text, the differences between the superconducting and normal-state solutions are considerably more pronounced at fixed momentum. The emergence of sharp coherence features in the occupied spectra at $k_F$ thus provides a spectroscopic signature of pairing, offering an experimentally accessible criterion for identifying superconducting correlations in ARPES measurements, even when the gap itself cannot be directly resolved.

\begin{figure}[t]{
    \centering
    \includegraphics[width=0.46\textwidth]{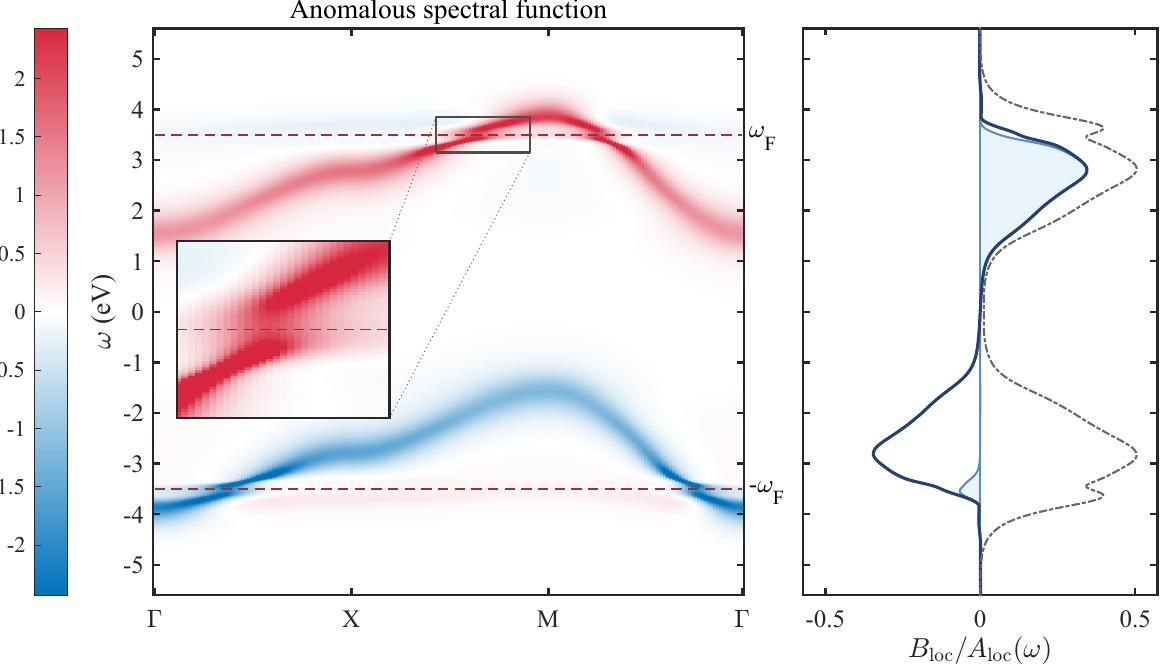}
\caption{Local and momentum-resolved anomalous spectra for the $\eta$-pairing superconducting state obtained from the TOA at $T^\text{eff}=1195\,\mathrm{K}$ and a photodoping level $\delta \approx 0.4$ ($\omega_F= 3.5\,\mathrm{eV}$).
Left panel: momentum-resolved anomalous spectral function for the square lattice along the high-symmetry path $\Gamma$--$X$--$M$--$\Gamma$.
Right panel: local anomalous spectral function $B_{\mathrm{loc}}(\omega)$ together with the corresponding occupation.
For comparison, the normal local spectral function $A_{\mathrm{loc}}(\omega)$ is overlaid as a gray dash-dotted line.
}
\label{fig_dispersion_offdiag}
}
\end{figure}

In addition to the occupied normal spectra discussed above, we also present the anomalous spectra extracted from the anomalous Green's function, which provide a complementary characterization of the $\eta$-pairing superconducting state.
Although the anomalous spectral function is not directly accessible in ARPES experiments, this spectrum serves as a useful reference for other theoretical studies. Figure~\ref{fig_dispersion_offdiag} shows the momentum- and frequency-resolved anomalous spectra, as well as the local anomalous spectral function $B_{\mathrm{loc}}(\omega)=-\mathrm{Im}\,G_{12}^{\mathrm{R}}(\omega)/\pi$, for the same parameters as in Fig.~2 of the main text.

In contrast to the normal spectral function, which is an even function of frequency with respect to $\omega=0$, the anomalous spectral function is odd in frequency and can therefore assume both positive and negative values.
As a consequence of the absence of positive definiteness, the local anomalous spectrum does not exhibit a clear gap near the effective Fermi level $\omega_F$ in the local spectral function. While the momentum-resolved anomalous spectra display gap-like features similar to those observed in the normal channel, there exists a faint band of incoherent spectral weight with opposite sign near $\omega_F$. Upon momentum integration, these contributions partially cancel the quasiparticle peak above the effective Fermi level, leading to a strongly suppressed secondary structure in the local anomalous spectrum.

Overall, the magnitude of the anomalous spectral function is comparable to that of the normal local spectrum, indicating that the system is close to the saturation regime of the superconducting order parameter. This demonstrates that the TOA is capable of accessing the ``deep'' $\eta$-pairing regime.

\section{Self-energy and quasiparticle dispersion in \texorpdfstring{$\eta$}{eta}-pairing superconducting states}

In this section, we present the real and imaginary parts of the self-energy corresponding to Fig.~3 of the main text over an extended frequency range, and explain how the quasiparticle dispersion and associated spectral weights follow from the self-energy in the $\eta$-pairing superconducting state. This analysis helps to understand how the self-energy controls the quasiparticle dispersion and lifetime.

\begin{figure}[t]
    \centering
    \includegraphics[width=0.46\textwidth]{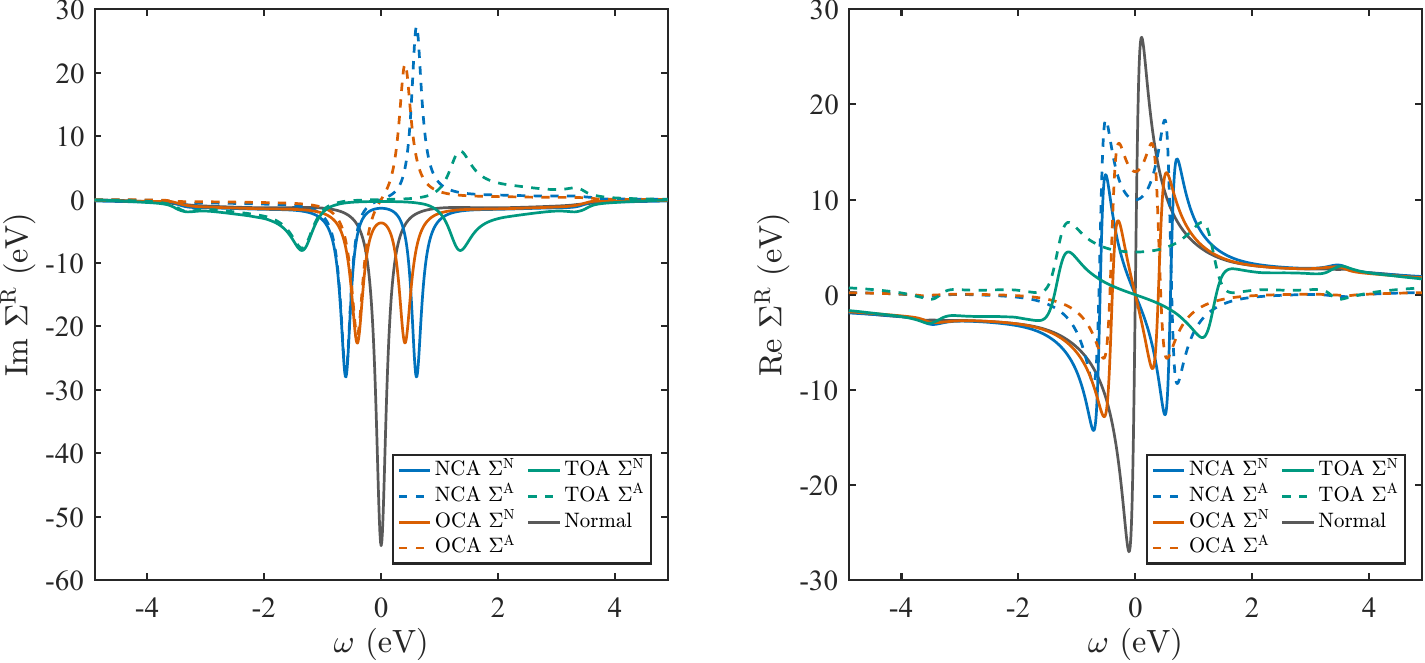}
    \caption{Imaginary (left panel) and real (right panel) parts of the retarded normal ($\Sigma^\mathrm{N}$) and anomalous ($\Sigma^\mathrm{A}$) self-energies for the NCA, OCA, and TOA cases shown in Fig.~3 of the main text, plotted over a broader frequency window. The parameters and notations are identical to those used in Fig.~3 of the main text.}
    \label{fig_selfenergy_largescale}
\end{figure}

In the main text (Fig.~3), we focused on the behavior of the imaginary part of the self-energy in the vicinity of the effective Fermi level and discussed the resulting quasiparticle dispersions. The self-energies over a broader frequency range were not shown there. For completeness, Fig.~\ref{fig_selfenergy_largescale} displays the real and imaginary parts of the retarded normal and anomalous self-energies for the same NCA, OCA, and TOA cases, now plotted over an extended frequency window. Since the retarded self-energy is analytic in the upper half of the complex plane, its real and imaginary parts are related through the Kramers--Kronig transformation.

For the normal self-energy in the normal state, the imaginary part exhibits a single-peak structure centered at $\omega=0$. Upon entering the $\eta$-pairing state, this peak splits into two, with the separation increasing as the order parameter grows. In the TOA results, the imaginary part already carries substantial weight in the upper and lower Hubbard bands, and an additional smaller feature appears near the Fermi level, corresponding to the frequency region magnified and analyzed in the main text.

From symmetry considerations, the imaginary part of the normal self-energy is an even function of frequency, while its real part is odd, as dictated by particle--hole symmetry. In contrast, for the anomalous self-energy, the imaginary part is odd and the real part is even, reflecting the combined effect of particle--hole symmetry and the imposed real symmetry-breaking seed.

We now turn to the derivation of the quasiparticle dispersion from the self-energy. As discussed previously, the lattice Green's function in the superconducting phase is represented in the Nambu basis as a $2\times2$ matrix. Owing to the symmetry of the $\eta$-pairing state and particle--hole symmetry, the inverse Green's function takes the form
\begin{equation}
\hat G^{-1}(\mathbf{k},\omega)=
\begin{pmatrix}
a(\mathbf{k},\omega) & b(\omega) \\
b(\omega) & a(\mathbf{k},\omega)
\end{pmatrix},
\label{eq:Ginv_simple}
\end{equation}
with
\begin{equation}
a(\mathbf{k},\omega)=\omega-\varepsilon_{\mathbf{k}}-\Sigma^{N}(\omega), \quad
b(\omega)=-\Sigma^{A}(\omega).
\end{equation}
Here $\Sigma^{N}$ and $\Sigma^{A}$ denote the normal and anomalous self-energies, respectively.
Both quantities are, in general, complex and frequency dependent.

The diagonal Green's function is obtained by inverting Eq.~\eqref{eq:Ginv_simple},
\begin{equation}
    \begin{aligned}
G_{11}(\mathbf{k},\omega)&=\frac{a(\mathbf{k},\omega)}{a(\mathbf{k},\omega)^2-b(\omega)^2}\\
&=\frac{1}{2}\left[
\frac{1}{a(\mathbf{k},\omega)-b(\omega)}
+
\frac{1}{a(\mathbf{k},\omega)+b(\omega)}
\right].
    \end{aligned}
\label{eq:G11}
\end{equation}
This identity shows that the diagonal Green's function is the sum of two effective propagators, corresponding to two quasiparticle branches.

In the normal state, the quasiparticle dispersions are defined by the condition
\begin{equation}
\omega-\varepsilon_{\mathbf{k}}-\mathrm{Re}\,\Sigma^{N}(\omega)=0.
\end{equation}
In the superconducting state described by Eq.~\eqref{eq:Ginv_simple}, this criterion can be generalized to the two branches appearing in Eq.~\eqref{eq:G11}.
Specifically, the quasiparticle energies $E_{\pm}(\mathbf{k})$ are defined by the real-part conditions
\begin{equation}
\omega-\varepsilon_{\mathbf{k}}-\mathrm{Re}\,\Sigma^{N}(\omega)
=\pm\,\mathrm{Re}\,\Sigma^{A}(\omega).
\label{eq:band_condition}
\end{equation}
Equation~\eqref{eq:band_condition} yields two quasiparticle branches. In the following, we analyze the spectral weight associated with these branches.

The diagonal spectral function is defined as
\begin{equation}
A_{11}(\mathbf{k},\omega)
=-\frac{1}{\pi}\,\mathrm{Im}\,G_{11}^{R}(\mathbf{k},\omega).
\end{equation}
To evaluate the spectral weight at the quasiparticle energies determined by Eq.~\eqref{eq:band_condition}, we write
\begin{equation}
a=a_1+i a_2, \quad b=b_1+i b_2,
\end{equation}
with
\begin{equation}
    \begin{aligned}
a_1=\omega-\varepsilon_{\mathbf{k}}-\mathrm{Re}\,\Sigma^{N}, &\quad
a_2=-\mathrm{Im}\,\Sigma^{N}, \quad\\
b_1=-\mathrm{Re}\,\Sigma^{A}, &\quad
b_2=-\mathrm{Im}\,\Sigma^{A}.
    \end{aligned}
\end{equation}

We first consider the ``$+$'' branch, for which the real-part condition $a_1=b_1$ holds.
In this case, the denominator of the first term in Eq.~\eqref{eq:G11} becomes purely imaginary,
\begin{equation}
a-b=i(a_2-b_2),
\end{equation}
leading to a dominant resonant contribution to the spectral function.
Evaluating the imaginary parts explicitly, one finds
\begin{equation}
A_{11}^{(+)}(\mathbf{k},\omega)\Big|_{a_1=b_1}
=
\frac{1}{2\pi}
\left[
\frac{1}{a_2-b_2}
+
\frac{a_2+b_2}{(2b_1)^2+(a_2+b_2)^2}
\right].
\label{eq:A11_plus}
\end{equation}
An analogous calculation for the ``$-$'' branch, defined by $a_1=-b_1$, yields
\begin{equation}
A_{11}^{(-)}(\mathbf{k},\omega)\Big|_{a_1=-b_1}
=
\frac{1}{2\pi}
\left[
\frac{1}{a_2+b_2}
+
\frac{a_2-b_2}{(2b_1)^2+(a_2-b_2)^2}
\right].
\label{eq:A11_minus}
\end{equation}

Equations~\eqref{eq:A11_plus} and~\eqref{eq:A11_minus} show that, at the quasiparticle energies defined by the real-part condition, the spectral peak intensity is primarily controlled by the first term, the imaginary parts of the combined self-energies $\Sigma^{N}\pm\Sigma^{A}$.
In particular, the dominant contribution to the peak height is given by
\begin{equation}
A_{11}^{\mathrm{peak}}(\mathbf{k})
\sim
\frac{1}{2\pi}\,
\frac{1}{-\mathrm{Im}\,[\Sigma^{N}(E_{\pm})\mp\Sigma^{A}(E_{\pm})]},
\label{eq:A11_peak}
\end{equation}
evaluated at energies $E_{\pm}$ satisfying
$\;E_{\pm}-\varepsilon_{\mathbf{k}}-\mathrm{Re}\,\Sigma^{N}(E_{\pm})
\pm\mathrm{Re}\,\Sigma^{A}(E_{\pm})=0$.
The second terms in Eqs.~\eqref{eq:A11_plus} and~\eqref{eq:A11_minus} provide subleading background corrections from the opposite branch.

The above analysis demonstrates that, within the simplified Green's function structure of Eq.~\eqref{eq:Ginv_simple}, the superconducting quasiparticle spectrum consists of two branches whose energies are determined by the real parts of $\Sigma^{N}\pm\Sigma^{A}$, while the corresponding spectral weights and lifetimes are governed by the imaginary parts of the same combinations.
This explains the main features of the momentum-resolved spectra.

\section{Optical conductivity}

Within the real-frequency formulation of steady-state DMFT, the optical conductivity can be evaluated directly as a function of real frequency~\cite{Li2020,Martin2008}. As a central experimentally accessible observable, the optical conductivity provides direct insight into charge dynamics, low-energy coherence, and collective transport properties of correlated electron systems.

The optical conductivity $\sigma_{\alpha\beta}(\omega)$ is related to the current--current response function
\(\chi_{\alpha\beta}(t,t')=\delta\langle j_{\alpha}(t)\rangle/\delta A_{\beta}(t')\) via~\cite{Martin2008}
\begin{equation}
    \sigma_{\alpha\beta}(\omega)=\frac{i}{\omega+i0^+}\chi_{\alpha\beta}(\omega).
    \label{sig_and_chi}
\end{equation}
The total susceptibility $\chi_{\alpha\beta}(\omega)$ consists of a paramagnetic contribution $\chi^{\mathrm{pm}}_{\alpha\beta}(\omega)$ and a frequency-independent diamagnetic term $\chi^{\mathrm{dia}}_{\alpha\beta}$~\cite{Martin2008}. In the normal state, these contributions cancel in the static limit,
\(\lim_{\omega\to 0}\chi^{\mathrm{pm}}(\omega)=-\chi^{\mathrm{dia}}\).
This cancellation does not occur in a superconducting state, which leads to a singular low-frequency response.

As a result, the optical conductivity can be decomposed into a regular part and a Drude contribution,
\begin{align}
    \mathrm{Re}\,\sigma(\omega) &= \pi D\,\delta(\omega) + \mathrm{Re}\,\sigma^{\mathrm{reg}}(\omega),\\
    \mathrm{Im}\,\sigma(\omega) &= \frac{D}{\omega} + \mathrm{Im}\,\sigma^{\mathrm{reg}}(\omega).
\end{align}
The real part $\mathrm{Re}\,\sigma(\omega)$ describes dissipative charge transport and is directly related to energy absorption from an external electromagnetic field, reflecting the spectrum of current-carrying excitations. The imaginary part $\mathrm{Im}\,\sigma(\omega)$ characterizes the reactive, nondissipative response associated with energy storage and long-lived coherent currents. The two components are related by Kramers--Kronig relations.

In a superconducting state, the appearance of a $\delta(\omega)$ contribution in $\mathrm{Re}\,\sigma(\omega)$ signals dissipationless dc transport, while the corresponding $D/\omega$ divergence in $\mathrm{Im}\,\sigma(\omega)$ reflects the inductive response of the superfluid. The Drude weight $D$ quantifies the phase stiffness of the superconducting condensate and serves as a hallmark of superconducting order.

In the following, we first focus on the regular part of the optical conductivity. In the absence of external fields and for an isotropic lattice system, vertex corrections to the optical conductivity vanish within single-site DMFT~\cite{Martin2008,Haule_2007}. This follows from symmetry considerations: the vertex correction term at frequency $\omega$ is proportional to
\(\sum_{\mathbf{k},\Omega} \mathbf{v}_{\mathbf{k}} G_{\mathbf{k}}(\Omega) G_{\mathbf{k}}(\Omega+\omega)\)~\cite{Haule_2007},
where $G_{\mathbf{k}}(\omega)$ denotes the single-particle Green's function and
\(\mathbf{v}_{\mathbf{k}} = \partial_{\mathbf{k}} \varepsilon_{\mathbf{k}}/\hbar\) is the current vertex.
For lattices with inversion symmetry, the Green's function is even in momentum, whereas the velocity is odd under $\mathbf{k}\rightarrow -\mathbf{k}$. Consequently, the momentum summation vanishes identically, and vertex corrections do not contribute.

The real part of the regular optical conductivity can therefore be computed solely from the bubble diagram as~\cite{Martin2008,Haule_2007}
\begin{equation}
\begin{aligned}
    \mathrm{Re}\,\sigma^{\mathrm{reg}}_{\alpha\beta}(\omega)
    =&\frac{e^2}{\hbar N_{\mathbf{k}}}
    \sum_{\mathbf{k}} v_{\mathbf{k}\alpha} v_{\mathbf{k}\beta}
    \int_{-\infty}^{\infty} \frac{d\omega'}{\pi\omega}
    \left[f(\omega')-f(\omega'+\omega)\right] \\
    &\times\mathrm{Tr}\left[\tau_3G_{\mathbf{k}}(\omega')\tau_3G_{\mathbf{k}}(\omega'+\omega)\right],
\end{aligned}
\label{sig_and_SF}
\end{equation}
where $\tau_3=\mathrm{diag}\{1,-1\}$ is a $2\times2$ matrix. 

Using Eq.~\eqref{sig_and_chi}, the imaginary part of the regular susceptibility follows as
\(\mathrm{Im}\,\chi^{\mathrm{reg}}_{\alpha\beta}(\omega)=\omega\,\mathrm{Re}\,\sigma^{\mathrm{reg}}_{\alpha\beta}(\omega)\). 
The real part of the regular susceptibility can then be obtained via the Kramers--Kronig relation:
\begin{equation}
\mathrm{Re}\,\chi^{\mathrm{reg}}_{\alpha\beta}(\omega)
=
\frac{2}{\pi}
\mathcal{P}
\int_{0}^{\infty}
\frac{\omega' \mathrm{Im}\,\chi^{\mathrm{reg}}_{\alpha\beta}(\omega')}
{\omega'^2-\omega^2}
\,d\omega'.
\label{eq:KK_chi}
\end{equation}

If the off-diagonal elements in $G_{\mathbf{k}}(\omega)$ are omitted in Eq.~\eqref{sig_and_SF}, the calculation yields a diagonal contribution $\mathrm{Re}\,\chi^{\mathrm{diag}}_{\alpha\beta}(\omega)$, which represents the response of the system in the absence of pairing-induced coherence. In the superconducting state, the emergence of anomalous correlations leads to a suppression of the finite-frequency spectral weight in the regular paramagnetic response. According to the Ferrell-Glover-Tinkham (FGT) sum rule~\cite{Ferrell1958,Tinkham1956}, this missing spectral weight is exactly compensated by the zero-frequency $\delta$-function. Consequently, the Drude weight $D$ can be extracted by evaluating the difference between the full paramagnetic response and its diagonal-only counterpart in the static limit:
\begin{equation}
D = \mathrm{Re}\,\chi^{\mathrm{reg}}_{\alpha\beta}(0) - \mathrm{Re}\,\chi^{\mathrm{diag}}_{\alpha\beta}(0),
\label{eq:Drude_extract}
\end{equation}
which allows for a consistent reconstruction of the full optical conductivity and the determination of the superfluid stiffness.

\section{Results for the cubic lattice and other lattice geometries}

In the main text, we focused on the photodoped two-dimensional square-lattice Hubbard model and argued that the key phenomena discussed there are not restricted to this specific geometry, but are expected to be generic for lattice systems in higher dimensions. To examine this point, we present additional results for other lattice geometries.

In particular, we consider the three-dimensional Hubbard model on the simple cubic lattice. DMFT is particularly well suited for higher spatial dimensions, where nonlocal correlations are reduced and the local self-energy description becomes more reliable~\cite{Metzner1989}. For the three-dimensional lattice, the noninteracting Hamiltonian yields the dispersion
\(
\varepsilon_{\mathbf{k}} = -2t \left(\cos k_x + \cos k_y + \cos k_z\right).
\)
When comparing different lattice dimensions, using the same hopping amplitude $t$ would result in substantially different bandwidths and energy scales.
To enable a meaningful comparison, we therefore fix $\sqrt{d}\,t$ (with $d$ the spatial dimension) to be constant across dimensions.
This choice ensures that the variance of the kinetic energy,
$N_{\mathbf{k}}^{-1}\sum_{\mathbf{k}} \varepsilon_{\mathbf{k}}^2$,
obtained by uniformly sampling the Brillouin zone, remains unchanged.
Such a scaling provides a natural normalization of the single-particle energy scale and is consistent with the standard procedure employed in DMFT~\cite{Metzner1989}.

\begin{figure}[t]{
    \centering
    \includegraphics[width=0.46\textwidth]{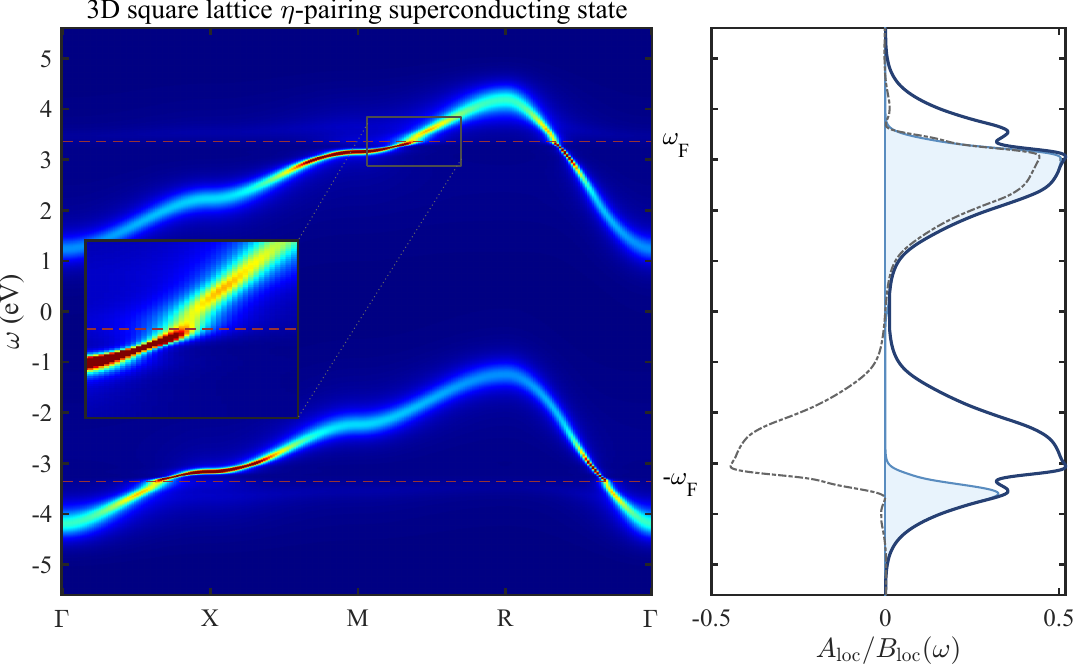}
\caption{Local and momentum-resolved spectra of the photodoped cubic-lattice Hubbard model obtained within DMFT using the TOA solver.
The calculation is performed at temperature $T^\text{eff}=1028\,\mathrm{K}$, below the superconducting transition temperature $T^\text{eff}_c=\mathrm{1427}\,\mathrm{K}$, for $\omega_F=3.36\,\mathrm{eV}$ and a photodoping concentration $\delta \approx 0.38$.
Left panel: momentum-resolved spectral function along the high-symmetry path $\Gamma$--$X$--$M$--$R$--$\Gamma$; the inset shows a magnified view near the Fermi energy.
Right panel: corresponding local normal spectral function and occupation.
For comparison, the local anomalous spectral function is overlaid as a gray dash-dotted line.
}
\label{fig_dispersion_3d}
}
\end{figure}

\begin{figure}[t]{
    \centering
    \includegraphics[width=0.46\textwidth]{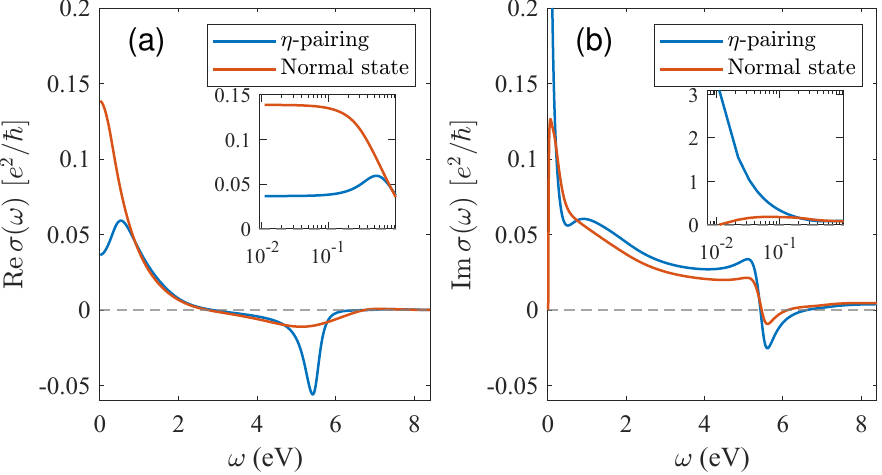}
\caption{Optical conductivity of the photodoped Hubbard model on the cubic lattice (same parameters as in Fig.~\ref{fig_dispersion_3d}).
(a) Real part of the optical conductivity, $\mathrm{Re}\,\sigma(\omega)$, excluding the zero-frequency $\delta$-peak contribution.
(b) Imaginary part of the optical conductivity, $\mathrm{Im}\,\sigma(\omega)$.
The low-frequency behavior is shown in the insets on a logarithmic frequency scale.
Blue lines correspond to the $\eta$-pairing superconducting state, while red lines show the normal-state results obtained by suppressing the anomalous component.
}
\label{fig_optical_conductivity_3d}
}
\end{figure}

In all calculations, the interaction strength is fixed to $U=5.6\,\mathrm{eV}$.
For the three-dimensional case at a photodoping concentration of $\delta \approx 0.38$ and an effective temperature of $T^\text{eff}=\mathrm{1028}\,\mathrm{K}$, we perform DMFT calculations using the TOA impurity solver.
The resulting momentum-resolved normal spectral function, together with the local normal and anomalous spectral functions, is shown in Fig.~\ref{fig_dispersion_3d}, while the corresponding real and imaginary parts of the optical conductivity are presented in Fig.~\ref{fig_optical_conductivity_3d}.

Despite the different lattice geometry and dispersion, the overall spectral features in three dimensions resemble those obtained for the two-dimensional square lattice. 
From a fit of the order parameter, we extract a transition temperature of $T^\text{eff}_c = 1427\,\mathrm{K}$, which is comparable to the two-dimensional result. 
The spectral functions also exhibit the opening of a superconducting gap in the three-dimensional case. 
Compared to the normal state, the real part of the optical conductivity is suppressed at low frequencies (although it does not become negative at the considered $T^\text{eff}$), while the imaginary part displays a characteristic $1/\omega$ behavior at low frequencies. These features qualitatively resemble those found for the two-dimensional model in the main text.

For completeness, we performed analogous DMFT calculations for the one-dimensional Hubbard chain. In this case, the superconducting transition temperature is strongly suppressed and no clear gap opening is observed. Since spatial correlations play a crucial role in one dimension, and the local DMFT approximation misses qualitative features of the correlated electronic structure, we do not present a detailed analysis of this case.

\begin{figure}[t]{
    \centering
    \includegraphics[width=0.4\textwidth]{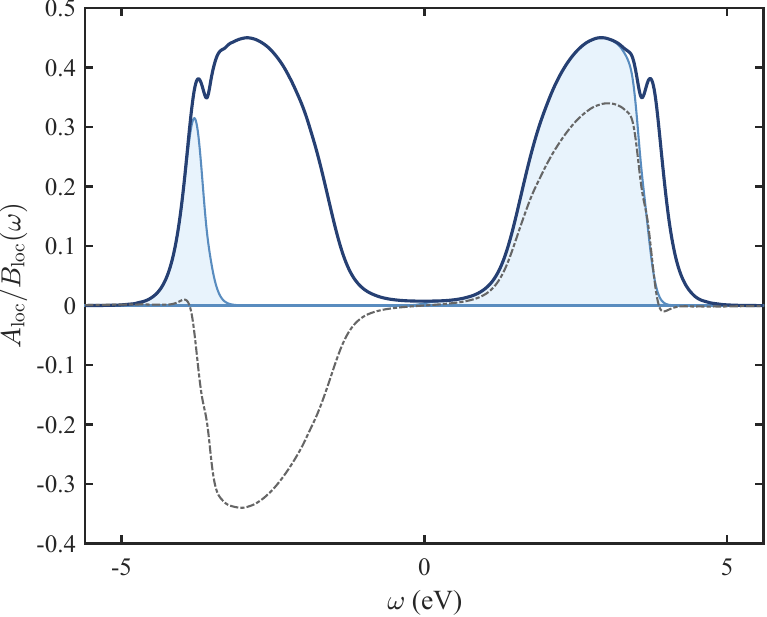}
\caption{
Local normal spectral function (dark blue line) and occupation (blue shading) for the Hubbard model on the infinitely-connected Bethe lattice, obtained within DMFT using the TOA solver. The calculation is performed at $T^\text{eff}=967\,\mathrm{K}$ and Fermi energy $\omega_F=3.64\,\mathrm{eV}$, corresponding to a photodoping concentration $\delta \approx 0.42$. For comparison, the local anomalous spectral function is overlaid as a gray dash-dotted line.
}
\label{fig_dispersion_bethe}
}
\end{figure}

We have further investigated the photodoped Hubbard model on the infinitely connected Bethe lattice, for which the DMFT solution becomes exact. The corresponding superconducting transition temperature is comparable to that obtained in two and three dimensions. Owing to the absence of translational symmetry on the Bethe lattice, momentum-resolved spectral functions are not accessible. Instead, Fig.~\ref{fig_dispersion_bethe} shows the local normal and anomalous spectral functions, as well as the occupation, at $T^\text{eff}=967\,\mathrm{K}$ and $\delta=0.42$. A pronounced suppression of spectral weight near the effective Fermi levels is visible in the local spectral function, indicating that the $\eta$-pairing state likewise develops a gap structure at low temperature.

Taken together, the results demonstrate that the high-temperature $\eta$-pairing superconducting state and its associated spectral gap are generic features of photodoped Mott insulators in two and higher dimensions.

\bibliography{mybibtex}